%% file: main.tex
\documentclass[preprint, review, 12pt]{elsarticle}

\usepackage{fullpage}
\usepackage{xcolor}

\usepackage{amssymb}

\usepackage{graphicx}
\usepackage[space]{grffile}
\usepackage{latexsym}
\usepackage{textcomp}
\usepackage{longtable}
\usepackage{multirow,booktabs}
\usepackage{amsfonts,amsmath,amssymb}
\usepackage{url}
\usepackage{hyperref}
\hypersetup{colorlinks=false,pdfborder={0 0 0}}
\usepackage[utf8x]{inputenc}
\usepackage[german,english]{babel}
\usepackage{bm} 
\usepackage[mathscr]{euscript} 
\usepackage{lipsum}
\usepackage{eurosym}
\usepackage{caption} 

\usepackage{comment} 
\usepackage{float} 
\usepackage{rotating} 
\usepackage{tablefootnote} 
\usepackage{scrextend} 
\usepackage{pbox} 
\usepackage{mathtools} 
\usepackage{pifont} 
\usepackage{csvsimple} 
\usepackage{pdflscape} 
\usepackage{gensymb} 
\usepackage{url} 
\usepackage[bottom]{footmisc} 
\usepackage{subfigure} 

\usepackage{nomencl}
\newlength{\nomitemorigsep}
\renewcommand\nomgroup[1]{%
\itemsep\nomitemorigsep
  \item[\bfseries
  \ifstrequal{#1}{A}{Acronyms and abbreviations}{%
  \ifstrequal{#1}{V}{Latin variables}{%
  \ifstrequal{#1}{G}{Greek variables}{%
  \ifstrequal{#1}{I}{Indices}{%
  \ifstrequal{#1}{P}{Superscripts}{%
  \ifstrequal{#1}{B}{Subscripts}{
  \ifstrequal{#1}{H}{Output variables of the Heat Infrastructure and Technology model}{%
  }}}}}}}%
]\itemsep\nomitemsep}
\makenomenclature

\usepackage{lineno}

\journal{Computers \& Chemical Engineering}

\usepackage{mathrsfs} 
\usepackage{booktabs}

\begin{document}

\begin{frontmatter}



\title{Detail or uncertainty? Applying global sensitivity analysis to strike a balance in energy system models}
\tnotetext[ESI]{Supplementary Material available.}


\author[1]{Maria Yliruka\corref{corresponding}}
\ead{m.yliruka@imperial.ac.uk}
\author[1]{Stefano Moret}
\author[1]{Nilay Shah}
\cortext[corresponding]{Corresponding author}

\address[1]{Department of Chemical Engineering, Imperial College London, Exhibition Road, London, SW7 2AZ,UK}

\begin{abstract}
Energy systems modellers often resort to simplified system representations and deterministic model formulations (\emph{i.e.}, not considering uncertainty) to preserve computational tractability. However, reduced levels of detail and neglected uncertainties can both lead to sub-optimal system designs. In this paper,  we present a novel method that quantitatively compares the impact of detail and uncertainty to guide model development and help prioritisation of the limited computational resources. By considering modelling choices as an additional `uncertain' parameter in a global sensitivity analysis, the method determines their qualitative ranking against conventional input parameters. As a case study, the method is applied to a peer-reviewed heat decarbonisation model for the United Kingdom with the objective of assessing the importance of spatial resolution. The results show that while for the optimal total system cost the impact of spatial resolution is negligible, it is the most important factor determining the capacities of electricity, gas and heat networks.

\textbf{Highlights}
\begin{itemize}
\item Quantitative method to define the appropriate level of detail in energy models
\item Global sensitivity analysis method to compare the importance of detail vs uncertainty
\item Data: uncertainty ranges are derived for key parameters of the UK energy system
\item Application to a heat decarbonization model to study the impact of spatial resolution
\item Spatial resolution does not impact total cost but is important for system design
\end{itemize}
\end{abstract}


\end{frontmatter}

\setlength{\abovedisplayskip}{0pt}

\textbf{Word Count:} 8481

\input{variables.tex}
\printnomenclature

\section{Introduction}
\label{sec:intro}
To meet the net-zero targets set by countries such as the United Kingdom (UK) \cite{hmgovernmentNetZeroStrategy2021}, the United States (US) \cite{unitedstatesdepartmentofstateandunitedstatesexecutiveofficeofthepresidentLongtermStrategyUnited2021} and members of the European Union (EU) \cite{croatiaandtheeuropeancommissionLongtermLowGreenhouse2020}, the power, heat and transport sectors need to be decarbonised by 2050. The large investments in energy technologies and infrastructure required in order to achieve this energy transition need to be resilient and economically viable under different scenarios of the future. Long-term energy system models (ESM) covering the time horizon from today until 2050 can be used to advise on the cost-optimal type, size and timing of these investments. To maintain computational tractability, these models often adopt simplified system representations, e.g. by reducing the temporal and spatial resolution of the input data or by neglecting operational detail. Additionally, most of these models are \textit{deterministic}, meaning that the uncertainties in input parameters like energy prices and demands are  not accounted for.\newline 
While simplified system representations were suitable to correctly size the technologies in fossil fuel based energy systems, a reduced level of temporal, operational and spatial detail can result in sub-optimal or even infeasible system designs when incorporating renewable power generation \cite{7350236,PONCELET2016631}. Specifically, temporal aggregation smooths the fluctuations in renewable power output and demand profiles, causing the baseload and renewable generation capacities to be oversized and the flexible generation and storage capacities to be undersized \cite{PONCELET2016631,PFENNINGER20171}. Similar errors in the capacity sizing are observed when neglecting operational constraints, particularly the need to maintain operating reserves \cite{ponceletLongtermEnergysystemOptimization2018}. The renewable generation capacity is overestimated as the operating reserves require thermal generators to be online even when renewables could meet the power demand. Furthermore, flexible generators are undersized because additional capacity of flexible generators are needed as non-spinning reserves given their fast start-up times and ramp rates. Spatial resolution is required to explicitly represent network connections and their capacities in ESM. The impact of spatial resolution is therefore dependent on whether the cost-optimal solution is constrained by network capacities \cite{FRYSZTACKI2021116726}: If the network is sufficiently sized or can be reinforced, the capacities of renewable technologies tend to be oversized at low spatial resolution because their capacity factors within aggregated areas are averaged out. If however the network is congested, the capacity of renewables can be undersized at low spatial resolution as the access to high capacity factor sites is restricted by the network and therefore higher capacities have to be build elsewhere. Overall, this shows that a reduced level of detail -- along the temporal, operational or spatial dimensions -- can lead to incorrect capacity sizing and total system cost evaluations in energy systems with high penetration rates of renewable energy. Additionally, operating costs and greenhouse gas (GHG) emissions are miscalculated as they vary strongly with the level of renewable energy integrated into the system.\newline Long-term ESM also inevitably rely on forecasts and assumptions to estimate input parameters when optimising the investment strategy. As recently highlighted by the reduction in energy demand during the SARS-CoV-2 pandemic and the subsequent increase in natural gas prices when economies restarted \cite{suzannahinsonEnergyPriceCrunch22}, these input parameters are subject to significant and unpredictable variations that can translate into changes to the cost-optimal system design. The total system cost was previously observed to be sensitive to economic parameters such as fuel prices, discount rates and capital costs of low-carbon power generators \cite{MORET2017597,MAVROMATIDIS2018219,PYE2015673,PILPOLA2020117652}. Optimisation under uncertainty allows to incorporate the uncertainty of input parameters into the model formulation. In particular, robust optimisation methods help to determine the least-cost system design which is feasible across all possible parameter values \cite{ben-talRobustOptimization2009}. In the case of the decarbonisation of the New York State power system, a multi-adaptive robust optimisation led to higher total capacities and a shift from onshore to offshore wind when accounting for the uncertainties of annual electricity demand, market availability of wind turbines and solar PV \cite{ZHAO2021100019}.\newline Unfortunately, both incorporating uncertainties and increasing the level of detail in ESM come at a price. As an example, dependent on the method used to construct the uncertainty set in \citet{ZHAO2021100019}, the solution time of the robust formulation increased by at least 887 times compared to the deterministic model. On the other hand, the computational time is generally observed to grow exponentially with increased temporal resolution; this also applies to spatial resolution  if networks between areas are explicitly modelled \cite{en12244656}, while if only the resolution of the renewable generation sites is modelled, a linear increase in computational time was reported \cite{FRYSZTACKI2021116726,PRIESMANN2019113783}. The increase in computational time caused by operational detail depends on the type of constraints that are introduced \cite{7038323}: constraints that only increase the model size come at a lower computational cost than constraints changing the type of the problem by introducing binary variables \cite{KOTZUR2021100063}.\newline Thus, energy system modellers face the decision of how to prioritise the limited computational resources given that both the level of detail and the uncertainty of input parameters impact the capacity investment strategy. Previous guidance has been qualitative, suggesting to start with a simple model formulation, subsequently increasing the level of detail and only considering uncertainty thereafter \cite{DECAROLIS2017184,KOTZUR2021100063}. However, adapting detailed, large-scale models for optimisation under uncertainty can be difficult \cite{PFENNINGER201474}. \citet{SCHYSKA20212606} proposed to quantify the differences in the optimal solution of scenarios due to different model structures or parameter values using a distance measure based on the objective function. Small distance measures indicate that the optimal solutions are part of each others' near-optimal solution space. The method is however limited by the pairwise comparison of linear programs and the inability to draw differentiated conclusions on different outputs other than the objective function value.\newline In this paper, we present the first method using a \textit{quantitative} analysis to rank the impact of detail against the impact of uncertainty for each output variable of interest. The fundamental idea underlying our method lies in including modelling choices, such as the level of detail, alongside uncertain techno-economic input parameters in a global sensitivity analysis (GSA). The ranking obtained as output of the GSA enables energy system modellers to tailor the model formulation to accurately answer the research question of interest by increasing the level of detail or including uncertainties. The novel GSA application is illustrated and verified using as a case study the impact of spatial resolution on a peer-reviewed urban heat decarbonisation model. For this case study, the uncertainty of techno-economic parameters was characterised using a large set of UK-specific data. The collected dataset and analysis can be useful to researchers and practitioners addressing the decarbonisation of the UK energy system.\newline The remainder of the paper is structured as follows: the literature review in Section 2 discusses how energy system modellers have previously selected the level of detail in long-term ESM. Section 3 describes the methodology, detailing how to include modelling choices in a conventional GSA. After the case study is introduced in Section 4, the results for the uncertainty characterisation, the impact of spatial resolution and its implications for model development are discussed in Section 5.

\section{Literature review} \label{Sect:LitReview}
Balancing detail, uncertainty and computational time becomes especially relevant for long-term ESM used in deep decarbonisation studies. In fact, on the one hand, the impact of detail and uncertainty increase with higher shares of renewable generation \cite{PFENNINGER20171} and longer time horizons \cite{MORET2017597}, respectively; on the other hand, the available computational resources are reduced by the need of extending the scope of ESM beyond power systems, including other sectors such as heat and transport. However, taking the 88 ESM for deep decarbonisation reviewed in \citet{BORASIO2022111730} as a sample, only 8 studies comment on the level of temporal and spatial detail used in their analysis. \citet{LOMBARDI20202185} explained their choice of a double-scale spatial representation based on data availability and administrative units with legislative power. The aggregation of hourly time series to time slices or representative days was in some cases justified by the limited computational time \cite{JAGEMANN2013622,SPIECKER2014185} or by referring to previous publications that had shown small deviations from full hourly models \cite{KRAKOWSKI2016501,hainschEmissionPathwaysLowCarbon2018a,en12152988}. But only \citet{PFENNINGER201583} and \citet{LIMPENS2019113729} explained the chosen level of temporal resolution by comparing the results of their simplified models to the system design and computational time obtained by running their models at hourly resolution. The review therefore suggests that justifying the choice of the level of detail or the use of a deterministic modelling approach has not yet become a standard procedure despite its impact on the optimal results.\newline Despite the lack of widespread application, separate methodologies have been proposed in the literature to assess either the impact of detail or the impact of uncertain input parameters.
The impact of detail was predominantly analysed using \emph{intramodel} comparisons, i.e. by first developing multiple formulations of a single modelling framework at different levels of detail and then analysing the differences in the model outputs across these versions. \emph{Intermodel} comparisons, on the other hand, use a soft-linking between at least two modelling frameworks by refining an initial capacity investment strategy, determined with a model at a low level of detail, using another model at higher resolution \cite{PONCELET2016631,DEANE2012303,kiviluomaComparisonFlexibilityOptions2018}.\newline
The impact of detail can be generally assessed using two different types of criteria: \emph{a priori} criteria, that are applied before the optimisation step by comparing the inputs, and \emph{a posteriori} criteria, that are based on the comparison of the optimisation results. The Root Mean Squared Errors between the original and aggregated time series for renewable power output \cite{NAHMMACHER2016430,KOTZUR2018474,HELISTO2020118585} and energy demand \cite{NAHMMACHER2016430,HELISTO2020118585,DOMINGUEZMUNOZ20113036,SCOTT2019113603,DESISTERNES2016368,VANDERHEIJDE201979,7527691} were used as \emph{a priori} criteria to assess temporal aggregation methods. The total system cost \cite{7350236,FRYSZTACKI2021116726,PRIESMANN2019113783,7741996,7038323,SIMOES2017183,HOFFMANN2021117825,PONCELET2020113843,HELISTO2021116712,FREW2016198}, the split in generation capacities \cite{7741996,HOFFMANN2021117825,PONCELET2020113843,HELISTO2021116712,nicolosi2010,PFENNINGER20171}, the amount of integrated renewable energy \cite{SIMOES2017183,PINA2013215}, the curtailment of renewables \cite{7350236,PONCELET2016631,FRYSZTACKI2021116726,FREW2016198,nicolosi2010} and the amount of GHG emissions \cite{7350236,PONCELET2020113843,7038323} were common \emph{a posteriori} criteria to quantify the impact of temporal, spatial and operational detail. The changes in total system cost were commonly small compared to changes in capacities \cite{7741996,SIMOES2017183}. By evaluating time aggregation methods using both types of criteria, \citet{HOFFMANN2021117825} however showed that \emph{a priori} criteria did not reliably predict the accuracy of the optimal results.\newline In the aforementioned studies, the `best' level of detail corresponds to the temporal and spatial resolution that provided the best approximation of the optimal results at full resolution with the least computational cost. Hence, the `best' level of detail is defined on a case-by-case basis as no quantitative criterion has yet been presented to generally identify the appropriate level of detail. \citet{NOLTING2022122522} summarised this trade-off between the improvement in accuracy and the cost connected with increased model detail in an abstract mathematical framework. Applying different temporal aggregation methods, both \citet{PONCELET2016631} and \citet{PFENNINGER20171} showed that the method applied to reduce the level of temporal detail can be more important than the number of representative days or time slices themselves. In the case of the operational detail, the linear relaxation of the full set of unit commitment constraints led to the best approximation of the optimal results determined for full operational detail in \citet{ponceletLongtermEnergysystemOptimization2018} and \citet{7038323}. This implies that, in contrast to the gradual increase of temporal or spatial resolution that allows to balance the computational cost with the accuracy of the result, the operational detail is best increased in a stepwise manner including all relaxed constraints at once. \newline In the literature, temporal, spatial and operational resolution are mostly analysed in isolation. However, in some studies their impacts are also compared to one another. The publications by \citet{PONCELET2016631,PONCELET2020113843} and \citet{HELISTO2021116712} agreed that the impact of operational detail is relatively small compared to temporal detail, especially for systems including battery storage or demand-side response. In \citet{FREW2016198}, the spatial resolution showed a higher impact on total system cost than the temporal resolution. Comparing the impact of all three modelling choices in terms of solving time, model size and objective function value, \citet{PRIESMANN2019113783} suggested to first reduce the operational, then the temporal and spatial details.\newline The impact of uncertain input parameters is often assessed \textit{via} sensitivity analysis. The method allocates the uncertainty in each output variable to the uncertainty of the different input parameters \cite{saltelliGlobalSensitivityAnalysis2008}. A global sensitivity analysis (GSA) considers all input parameters simultaneously, whereas a local sensitivity analysis (LSA) only studies the impact of a selected set of input parameters in a \emph{one-at-a-time} fashion. The impact of each input parameter on one or more outputs of interest can be quantified by variance-based GSA methods. However, for ESM, variance-based methods are generally computationally expensive because of the large number of inputs \cite{saltelliSensitivityAnalysisChemical2005,KUCHERENKO20091135}. In these cases, a qualitative ranking of the input parameters can be obtained by applying the computationally efficient Elementary Effects (EE) method, first proposed by \citet{doi:10.1080/00401706.1991.10484804}, which does not require calculating the variance of the output. \citet{MORET2017597} showed that an uncertainty characterisation should precede the GSA because a false ranking was obtained when using generic uncertainty ranges. GSA was applied to urban \cite{LYTHCKEJORGENSEN201634,MAVROMATIDIS2018219,PETKOV2020115197} and national \cite{PYE2015673,MORET2017597,PILPOLA2020117652,FAIS2016154} ESM, in some cases focusing on highly renewable systems \cite{PIZARROALONSO2019113528}. In all these studies, the impact of uncertain energy demand, resource availability and techno-economic parameters such as capital costs, efficiencies and fuel prices was only quantified at a single level of model detail. Similarly, only a few of the publications analysing the impact of the level of detail apply a LSA. As an example, \citet{SIMOES2017183} study the impact of spatial resolution and also vary the investment and operation \& maintenance costs of wind technologies. \citet{JALILVEGA2018339} also focus on spatial detail and, additionally, consider variations in the electricity, gas and carbon prices. \citet{PONCELET2020113843} performed a LSA to study the impact of operational detail under different assumptions on the cycling characteristics of thermal power generators and operating reserve requirements. However, these last three works only evaluated whether the impact of the level of detail was still significant under different sets of input data, but did not comparatively assess the importance of detail against the uncertainty of the model inputs.\newline Our review shows that a method allowing to quantitavely compare the impact of modelling choices such as the level of detail and the uncertainty of input parameters is still missing in the literature. Hence, the novel GSA application presented in the following Section fills the gap identified in the literature by providing a methodological framework to systematically guide model development. 

\section{Methodology}
\begin{figure*}
\centering
  \includegraphics[width=17.1cm]{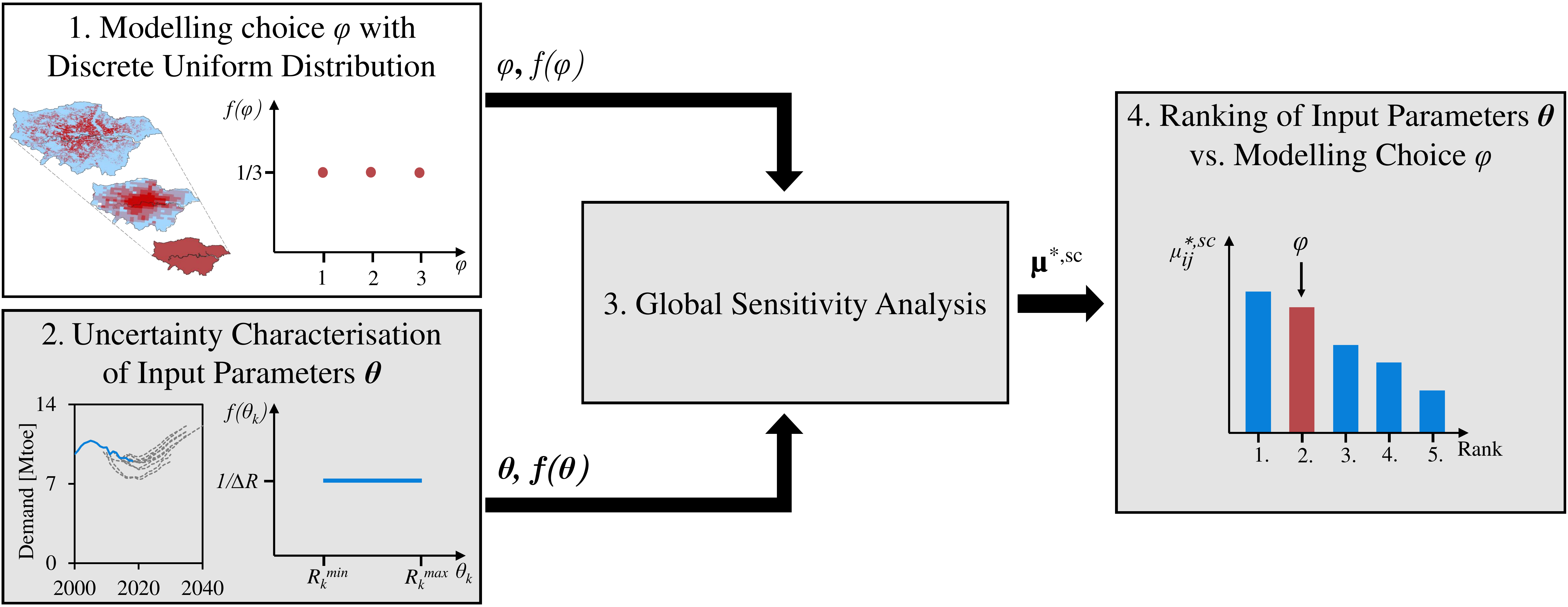}
  \caption{Summary of the main steps for the proposed GSA procedure. The novel step (1) is highlighted in white while a grey background is used for the three steps (2, 3, 4) of the conventional GSA. The modelling choice $\varphi$ is considered as an additional `uncertain' input with a discrete uniform distribution $f(\varphi)$ (1) alongside conventional input parameters $\boldsymbol{\mathrm{\theta}}$, each $\theta_k$ characterised by a uniform distribution $f_k(\theta_k)$ (2). In the GSA step, the input space is sampled based on $f(\varphi)$ and $\boldsymbol{f\mathscr{(\theta)}}$ (3). Finally, a qualitative ranking of $\varphi$ against $\boldsymbol{\mathrm{\theta}}$ for each output of interest $y_j$ is derived based on the sensitivity measure $\mu_{ij}^{*,sc}$ (4).}
  \label{fgr:Method}
\end{figure*}

The trade-off between modelling choices and uncertainty of input parameters can be assessed by using GSA methods in a novel way. Following the categorisation of uncertainties discussed by \citet{KIUREGHIAN2009105}, modelling choices can be seen as \textit{epistemic} uncertainties, i.e. uncertainties that can be reduced by increasing the level of model detail, in contrast with the \textit{aleatory} uncertainty of parameters like energy demands, fuel prices or technology learning rates, which cannot be reduced. A conventional GSA needs as inputs the uncertainty distributions $\boldsymbol{f\mathscr{(\theta)}} = \left[ f_1(\theta_1),...,f_K(\theta_K) \right]^T$ of the $K$ input parameters $\boldsymbol{\mathrm{\theta}} = \left[ \theta_1, ..., \theta_K \right]^T$. Modelling choices are not commonly referred to as `input parameters' and were therefore excluded in the GSA studies reviewed in Section 2. The novelty of our approach lies in including the modelling choice of interest as an additional `uncertain' parameter $\varphi$, characterised by a discrete uniform distribution $f(\mathrm{\varphi})$. In this way, the relative impact of conventional input parameters and modelling choices can be determined in a single methodological framework. For clarity, the term \textit{input factors} will be used in the following to collectively refer to the modelling choice $\varphi$ and input parameters $\boldsymbol{\mathrm{\theta}}$. The vector of input factors is defined as $\mathbf{x} = [\theta_1, ..., \theta_K, \varphi]^T$, where $x_i$ refers to an element of $\mathbf{x}$.
The modified GSA method is illustrated in Fig. \ref{fgr:Method} and comprises three main steps:
\begin{enumerate}
    \item Translating each modelling choice $\varphi$ into a discrete uniform distribution;
    \item Characterising the uncertainty of conventional input parameters $\boldsymbol{\mathrm{\theta}}$;
    \item Applying a GSA to comparatively rank the input factors $\boldsymbol{\mathrm{x}}$.
\end{enumerate}
These steps are explained in more detail in the following subsections.
The method is described such that it can be used in any GSA software package. For our work, the MATLAB scripts for factor sampling and EE-based sensitivity measures developed by the University of Florida were adapted \cite{yogeshkhareandrafaelmunoz-carpenaMorrisSUSamplingn.d.}.
\subsection{Translating each modelling choice into a discrete uniform distribution}\label{Sect:step1}
In this step, $p_{\varphi}$ different model formulations must be developed, that only differ in a single modelling choice. In the discrete uniform distribution, each level of detail $p_{\varphi}$ has an equal probability of 1/$p_{\varphi}$.\newline As an example, let us consider a power system model where the impact of temporal resolution  on the capacity deployment of onshore wind is of interest. The original time series could be aggregated into 1, 4, 6 and 12 representative days. Therefore, $p_{\varphi}=4$ different model formulations must be developed in this case. The associated discrete uniform distribution has four discrete levels, each with probability $1/4$. \newline The variation of the operational detail may introduce additional parameters such as ramp rates or minimum load points. In this case, the impact can only be compared for input parameters shared across all model formulations.

\subsection{Characterising the uncertainty of conventional input parameters}\label{Sect:step2}
An uncertainty distribution $f_k(\theta_k)$ for each input parameter $\theta_k$ (with $k=1,...,K$) of the ESM must be developed based on the collected data. The uncertainty characterisation was based on the methodology presented in \citet{MORET2017597}. In \ref{Sct:AppendixA}, the detailed processing of the individual UK data sources is described.\newline External models, error in forecasts, historical and manufacturer data were used and could in general be complemented by a collection of values reported in the literature and expert opinions. Inconsistencies across a series of reports led to an exclusion of the corresponding data.\newline Continuous uniform distributions were used for all parameters because the limited data did not suggest otherwise. For a conservative estimate of the uncertainty, the lower and upper bounds ($R^{min}_k$, $R^{max}_k$) for each input parameter $\theta_k$ were defined based on the extreme values identified across the different sources. In cases where the techno-economic parameter $\theta_k$ was a function of the capacity or, in the case of heat pumps, of the inlet temperature, $R^{min}_k$ and $R^{max}_k$ were determined based on the fit function instead. The final relative ranges $R^{\%,z}_k$ with $z \in \{min,max\}$ were calculated with reference to the nominal value $R^0_k$ as follows:
\begin{equation}
    R^{\%,z}_k=\frac{R^{z}_k-R^0_k}{R^0_k}, \quad  R^0_k \neq 0
\end{equation}
Limited by the data availability, the forecast errors and the applicability of historical values were only assessed over a 5- and 10-year time interval, respectively. In the case study, the bounds $R^{\%,min}_k$ and $R^{\%,max}_k$ on each input parameter $\theta_k$ were assumed to remain constant throughout the 30-year time horizon except for the uncertainty ranges of the energy demands. As the forecast error for energy demands in the US was previously observed to increase across a 15-year time horizon \cite{MORET2017597}, the ranges derived based on errors in UK forecasts were linearly extrapolated from 5 to 15 years and thereafter kept constant.

\subsection{Applying a GSA to rank the input factors}
To obtain a ranking of a given modelling choice $\varphi$ against $K$ conventional input parameters $\theta_k$, $k=1,...,K$, the input factor space needs to be sampled first to subsequently calculate the EE with respect to the output variables of interest. To identify correlations among input factors, $G$ sample points $\boldsymbol{\mathrm{\gamma}} = [\gamma_1, ..., \gamma_G]^T$ should be spread across the entire input factor space. As the $I$ input factors, with $I=K+1$ comprising $K$ input parameters $\theta_k$ and one modelling choice $\varphi$, can have different order of magnitudes, they are scaled by drawing samples from a $I$-dimensional unit hypercube first that are then projected onto the actual user-defined distributions developed in the steps outlined in Section \ref{Sect:step1} and \ref{Sect:step2}.\newline Input factors are varied in a  \emph{one-at-a-time} fashion along $T$ trajectories, with each trajectory $t=1,...,T$ starting from a different point in the unit hypercube. Hence, $G=T(I+1)$ sample points are generated for $I$ input factors. The input space is discretised into a $p_{\theta}$-level grid, changing each parameter $\theta_k$ by $\Delta_{\theta}$ with $\Delta_{\theta}=p_{\theta}/2(p_{\theta}-1)$. The only exception is the modelling choice $\varphi$, which is already discretised into $p_{\varphi}$ levels based on the $p_{\varphi}$ different model formulations developed in Section \ref{Sect:step1}, and is therefore varied by $\Delta_{\varphi}$ with $\Delta_{\varphi}=1/(p_{\varphi}-1)$ instead.\newline The enhanced Sampling for Uniformity (eSU) method \cite{chitaleEffectiveParameterScreening2017} was applied which evenly samples all $p_{\theta}$ and $p_{\varphi}$ levels and spreads the $T$ trajectories across the full input space. It outperforms methods relying on oversampling the input space to select $T$ trajectories with the maximum Euclidean distance in terms of computational time and sample spread \cite{chitaleEffectiveParameterScreening2017}.\newline Overall, the model needs to be computed $G=T(I+1)$ times to calculate $T$ $EE_{ij}(t)$ for each input factor $x_i$, $i = 1, ..., I$ with respect to each output variable $y_j$ with $j=1, ...,J$. For each trajectory \emph{t}, $EE_{ij}(t)$ is calculated as the ratio between the change in the input factor $x_i$ and the consequent change in the output variable, $\delta y_{ji}(t)$. The EE for the $K=I-1$ conventional input parameters $\theta_k$, $EE_{ij}(t)$ is defined as 

\begin{equation}
    EE_{kj}(t) = \frac{\delta y_{jk}(t)}{\Delta_{\theta}}, \quad k=1,...,I-1
\end{equation}

for $t = 1, ..., T$ while for the modelling choice $\varphi$ as the $I$-th input factor 

\begin{equation}
    EE_{Ij}(t) = \frac{\delta y_{jI}(t)}{\Delta_{\varphi}}.
\end{equation}

In the original method proposed by Morris, the global sensitivity measures were based on the mean ($\mu_{ij}$) and standard deviation ($\sigma_{ij}$) of the $EE_{ij}(t)$ distribution \cite{doi:10.1080/00401706.1991.10484804}. But for non-monotonic models, positive and negative $EE_{ij}(t)$ values might cancel each other out leading to a negligible $\mu_{ij}$. Therefore, \citet{CAMPOLONGO20071509} introduced $\mu_{ij}^*$, the mean of the absolute values of $EE_{ij}(t)$, and showed empirically that it is a good proxy for the total sensitivity index determined using computationally expensive, variance-based methods: 

\begin{equation}
    \mu_{ij}^* = \frac{1}{T}\sum_{t=1}^{T} \left| EE_{ij}(t) \right|
\end{equation}

As we want to compare the impact on multiple output variables $y_j$ of varying magnitudes and units, $\mu_{ij}^*$ needs to be scaled. Using a similar approach to \citet{SIN2009925}, $\Tilde{\sigma}_j$, the standard deviation  of the distribution of $y_j(g)$ with $g=1,...,G$ over all $G$ sample points $\gamma_g$, is used to calculate $\mu_{ij}^{*,sc}$: 

\begin{equation}\label{eq:scaling}
    \mu_{ij}^{*,sc} = \frac{1}{\Tilde{\sigma}_j}\mu_{ij}^*
\end{equation}

Finally, the input factors can be ranked according to the magnitude of $\mu_{ij}^{*,sc}$. In this way, the impact of modelling choices and input parameters can be compared for design and operational variables. While $\mu_{ij}^{*,sc}$ is a good enough estimate to identify the correct ranking, the absolute impact of selected input factors should separately be quantified by variance-based methods, if needed.

\section{Case study}\label{Sct:HITmodel}
\begin{figure}[h]
\centering
  \includegraphics[height=6cm]{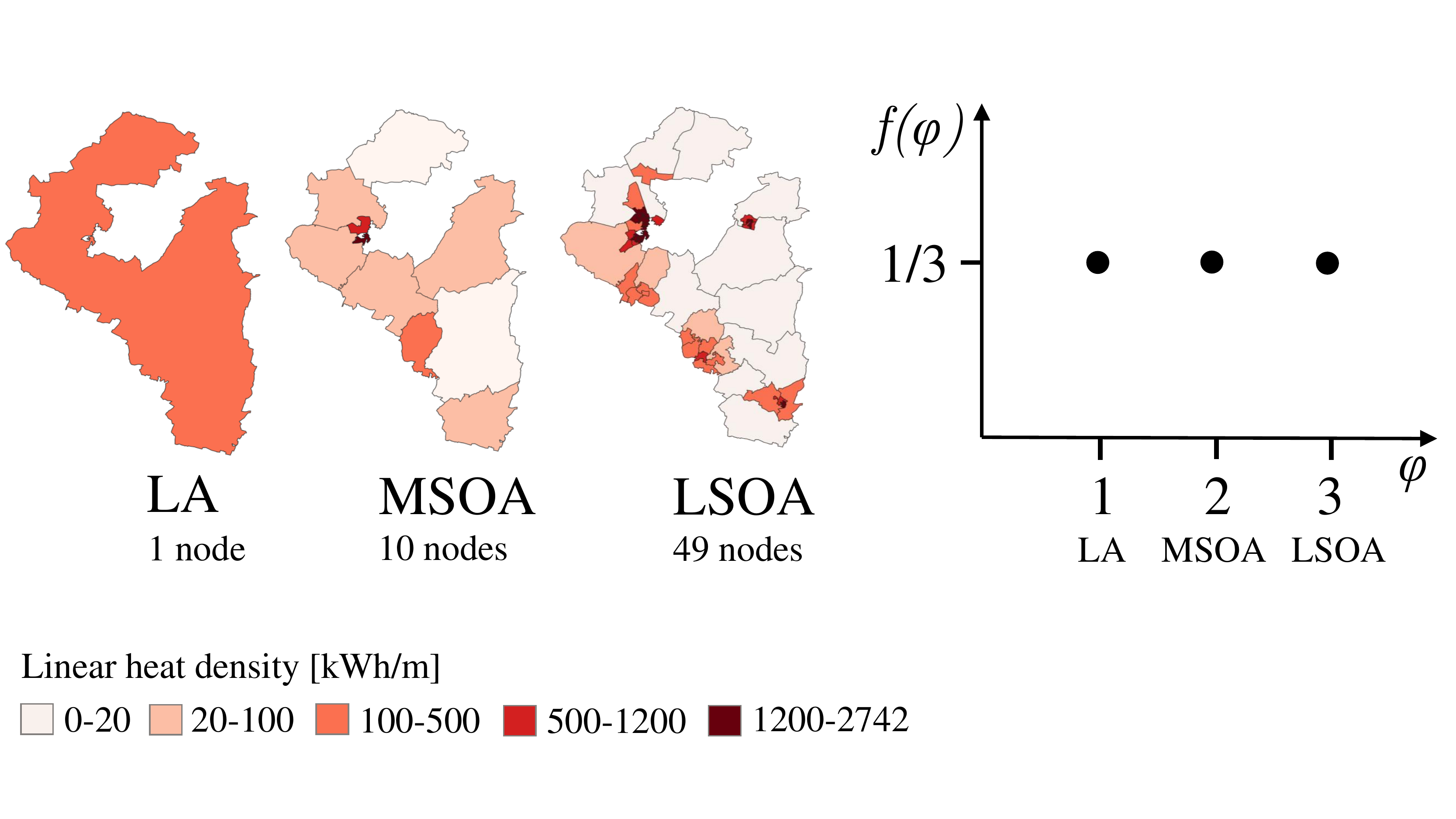}
  \caption{The heat demand distribution within the local authority of Winchester is shown for three levels of spatial resolution (in ascending order of detail, LA, MSOA and LSOA). To compare the impact of spatial resolution to the uncertainty of input parameters, the modelling choice $\varphi$ is translated into a discrete uniform distribution $f(\varphi)$ with equal probability $1/3$ for all three levels.}
  \label{fgr:Winchester}
\end{figure}

The novel GSA method is applied to the peer-reviewed Heat Infrastructure and Technology (HIT) model \cite{JALILVEGA20181051} to measure the impact of spatial resolution against the impact of techno-economic uncertain input parameters. In this heat supply model, a spatially explicit demand representation is adopted to analyse the trade-off between gas, electricity, and district heating (DH) for domestic heat decarbonisation by 2050. Using a mixed-integer linear formulation, the HIT model determines the design and operation strategies by minimising the total system cost. The full model formulation can be found in the Appendix of Jalil-Vega and Hawkes \cite{JALILVEGA20181051}. Air-source heat pumps (ASHP), ground-source heat pumps (GSHP), natural gas boilers (B) and gas-fired combined heat and power (CHP) units at different capacities can be deployed as individual or district heating (DH) technologies supplying heat exchangers (HEX). Additionally, electric resistance radiators (Erad) and solar photovoltaic (PV) are included as individual supply technologies. Based on the data availability for domestic electricity and gas demand in the UK, the HIT model can be formulated for Local Authorities (LA) as a whole, middle layer super output areas (MSOA) and lower layer super output areas (LSOA). Super output areas correspond to geographical entities at which Census data is provided by the Office for National Statistics \cite{officefornationalstatisticsCensusGeographyn.d.}. In the HIT model, the term \emph{node} refers to an entity into which the geographical region is subdivided.\newline The inter- and intranodal electricity, gas and heat networks are explicitly modelled. The $\emph{internodal}$ network capacity ($\emph{ICN}$) connects the centroids of two MSOA or LSOA along their linear distance. These internodal networks are required for the distribution of electricity and gas to each MSOA and LSOA as the area is only connected to the transmission grid in its northern and southern extremities. \emph{Intranodal} networks ($\emph{NLN}$) are assumed to be built along the road network within each node. The network length thereby scales linearly with the share of the peak heat demand met by electricity, gas or district heating. The initial internodal and intranodal gas and electricity networks are sized according to the initial domestic demands in 2015. These networks need to be reinforced as the lifetime of the initial pipelines is limited and additional capacity is required for the electrification of heat demands. In contrast, DH networks are new builds.\newline Applying the HIT model to six LAs with varying rural-urban character, the impact of spatial detail had previously been observed to change the DH network uptake \cite{JALILVEGA2018339}. In that work, given the heterogeneous distribution of heat demands as shown in Fig. \ref{fgr:Winchester}, the LA of Winchester showed the strongest spatial dependence and is therefore chosen as a case study to validate our method. The modelling choice of 1, 10 and 49 nodes is translated to the discrete uniform distribution shown in Fig. 
\ref{fgr:Winchester}. Because the input factors are scaled during the GSA, generic values $\{1,2,3\}$ instead of the actual number of nodes $\{1,10,49\}$ can be used to indicate the levels of spatial resolution $\varphi$.

With the exception of the lifetime of technologies and networks, all HIT model parameters are considered as uncertain: fuel prices ($\emph{Cost}^E$/$\emph{Cost}^G$, where $E$ indicates electricity and $G$ indicates natural gas), capital cost of technologies ($\emph{Cost}^C_\emph{tech}$), technological performance ($\eta^{\emph{Th}}_\emph{tech}$ /$\emph{COP}_\emph{tech}$/$\eta^E_\emph{tech}$, where \emph{Th} indicates thermal energy), capital cost of intranodal network capacity ($\emph{Cost}^{ND}_{E/G/H}$, where $H$ indicates heat), capital cost of internodal network capacity ($\emph{Cost}^{NT}_{E/G/H}$), operation \& maintenance cost ($\emph{Cost}^M_\emph{tech}$), losses in heat networks ($\emph{Loss}_H$), discount rate ($r$), electricity and heat demand ($\emph{Dem}^E$/$\emph{Dem}^H$).\newline For the factor sampling and sensitivity measure calculations MATLAB 2020a \cite{MATLAB2020a2020} was used. The HIT model formulated in GAMS 30.1 \cite{GeneralAlgebraicModeling2020} was solved with CPLEX V12.10.0 \cite{ILOGCPLEXOptimization21} on Intel Xeon E52667 machines (3.2 GHz, 8 cores, 16 logical processors, 128 GB RAM) with an optimality gap of 1\% by assigning 16 cores.

\section{Results and discussion}
\subsection{Uncertainty characterisation}
The uncertainty ranges determined based on UK-specific data sources are summarised in Table \ref{tbl:UncertaintyRanges_UK} and briefly discussed in this Section. In \ref{Sct:AppendixA}, a comprehensive discussion on the uncertainty ranges and the processing of the original data can be found. Also the data points and the respective uncertainty ranges are summarised in form of tables and figures in \ref{Sct:AppendixA}. The full data set is available in the Supplementary Material. The different magnitudes of the uncertainty ranges underline the value of an uncertainty characterisation because assuming the same relative ranges for all input parameters would have likely resulted in over- or underestimations.
\begin{table}[H]
\small
    \caption{\ Summary of the uncertainty ranges for the input parameters based on UK-specific data. The references and analysis can be found in the \ref{Sct:AppendixA}.}
    \label{tbl:UncertaintyRanges_UK}
    \begin{tabular*}{0.48\textwidth}{@{\extracolsep{\fill}}llll}
    \cline{1-4}
    Input parameter $\theta_k$ & $R_k^{\%,min}$ [$\%$] & $R_k^{\%,max}$ [$\%$] & Section\\
    \cline{1-4}
    $Dem^E$ & 0 & 15 & \multirow{2}{*}{\ref{Sct:AppDemand}} \\
    $Dem^H$ & -40 & 6 & \\
    $Cost^{CO2}$ & -50 & 50 & \ref{Sct:AppCO2}\\
    $Cost^E$ & -12 & 15 & \multirow{2}{*}{\ref{Sct:AppFuelPrices}}\\
    $Cost^G$ & -38 & 0 & \\
    $r$ & -81 & 0 & \ref{Sct:AppDiscRate}\\
    $COP_{ASHP}$ & -14 & 12 & \multirow{2}{*}{\ref{Sct:AppASHP}} \\
    $Cost^C_{ASHP}$ & -31 & 49 &  \\
    $COP_{GSHP}$ & -12 & 12 & \multirow{2}{*}{\ref{Sct:AppGSHP}} \\
    $Cost^C_{GSHP}$ & -40 & 44 & \\
    $\eta^{E}_{CHP}$ & -10 & 11 & \multirow{2}{*}{\ref{Sct:AppCHP}} \\
    $Cost^C_{CHP}$ & -11 & 9 & \\
    $\eta^{E}_{PV}$ & -14 & 17 & \multirow{2}{*}{\ref{Sct:AppPV}}\\
    $Cost^C_{PV}$ & -57 & 76 & \\
    $\eta^{Th}_{B/Erad/HEX}$ & -0.2 & 0.4 & \multirow{2}{*}{\ref{Sct:AppBoiler}}\\
    $Cost^C_{B/Erad/HEX}$ & -40 & 42 & \\
    \cline{1-4}
    \end{tabular*}
\end{table}
Specific data for the capital costs or efficiencies of electric resistance heaters and heat exchangers was not available. Compared to heat pumps, less design variations were expected for electric resistance heaters or plate heat exchangers commonly used in DH systems \cite{skagestadDistrictHeatingCooling1999}. Therefore, the uncertainty ranges for electric resistance heaters ($\eta^{Th}_{Erad}$, $Cost^C_{Erad}$) and heat exchangers ($\eta^{Th}_{HEX}$, $Cost^C_{HEX}$) in DH networks were assumed to be equal to the ranges derived for natural gas boilers. The ranges of network costs ($Cost^{NT}_{E/G/H}$,$Cost^{ND}_{E/G/H}$), losses ($\emph{Loss}_H$) and operation \& maintenance cost ($Cost^{M}_{tech}$) in Table \ref{tbl:UncertaintyRanges_CH} were taken from \cite{MORET2017597} because UK-specific data was not available.
\begin{table}[H]
\small
    \caption{\ Uncertainty ranges identified by \citet{MORET2017597} used for input parameters for which UK-specific data was unavailable }
    \label{tbl:UncertaintyRanges_CH}
    \begin{tabular*}{0.48\textwidth}{@{\extracolsep{\fill}}lll}
    \hline
    Input parameter $\theta_k$ & $R_k^{\%,min}$ [$\%$] & $R_k^{\%,max}$ [$\%$] \\
    \hline
    $Cost^{NT}_{E/G/H}$ & -39 & 39  \\
    $Cost^{ND}_{E/G/H}$ &  -39 & 39\\
    $Loss_H$ & -2 & 2 \\
    $Cost^{M}_{tech}$ & -48 & 36 \\
    \hline
    \end{tabular*}
\end{table}
Given that 85\% of the UK households use a gas boiler \cite{committeeonclimatechangeNextStepsUK2016}, the uncertainty in domestic natural gas demand was taken as a proxy for the uncertainty in domestic heat demand. The uncertainty is larger than for domestic electricity demand because the majority of natural gas is consumed for space heating, which is weather dependent.\newline For the technologies, the uncertainty ranges for efficiencies and coefficient of performance (COP) are less than half of the uncertainty ranges for the investment cost. In fact, although the efficiencies and COPs can vary with the design of the technology, they are fundamentally limited by the laws of thermodynamics. In contrast, the investment cost and fuel prices are dependent on market structures and global events, and hence they are subject to significant variations unless regulated. The comparison additionally suggests that estimating the thermodynamic performance of a future energy system might be more accurate than the evaluation of its total cost.\newline The uncertainty of efficiency or COP is similar across the technologies except for natural gas boilers. Here, the uncertainty range is smaller by more than an order of magnitude probably because the governmental legislation first enforced the use of condensing boilers \cite{officeofthedeputyprimeministerBuildingAct19842005} and more recently set a minimum efficiency of 92\% for newly installed boilers in English homes \cite{departmentforbusinessenergyindustrialstrategyHeatBuildingsBoiler2017}.\newline The relative ranges for the specific cost of ASHP, GSHP and boilers fall in the same order of magnitude. For monocrystalline solar PV, this range is exceeded by more than a factor of 1.5. In \cite{andreas_v_olympios_2021_4692649}, which was used to characterize the uncertainty of PV prices, no additional information regarding the manufacturer or the assembly of the solar PV panels is provided. Therefore, the reason for the larger uncertainty range could not be inferred. The uncertainty range for the specific cost of CHPs is comparably small. As the economies of scale \cite{haldiEconomiesScaleIndustrial1967} apply to the specific cost of the CHPs, a power law was fitted to all available of data points. The spread of the data points across a large capacity interval led to small uncertainty ranges compared to the other technologies.\newline While historical wholesale prices for electricity and natural gas are available, the uncertainty range for the carbon price is based on the sensitivities assumed in UK government reports. The uncertainty ranges for wholesale fuel prices are based on the decade between 2010 and 2020 therefore excluding the price increase since August 2021 \cite{WhatRocketingNaturalgas2021}. On an annual average, the wholesale natural gas prices declined between 2010-2019. However, if wholesale prices in 2021 were also included in the data set, the upper bound $R^{\%,max}$ would increase to 106\% for natural gas and 164\% for electricity. Overall, the uncertainty ranges for the fuel prices are comparatively small to the range of [-69.9\%, 232\%] for natural gas price forecasts in the US between 1984-2014 \cite{MORET2017597}.  
This difference can be explained by the fact that \citet{MORET2017597} analysed wellhead gas prices while the ranges calculated here are based on wholesale prices. As wholesale prices include additional costs such as network costs and taxes, they are arguably exposed to smaller fluctuations in relative values.\newline For the interest rate, the uncertainty range was calculated by comparing the annual average of the historical and actual interest rates of long-term UK government bonds from 1998 until 2020. The uncertainty range is strongly asymmetric because the long-term nominal rates for OECD economies such as the UK, the US and Japan have steadily declined since 1995 \cite{executiveofficeofthepresidentoftheunitedstatesLongtermInterestRates2015}. Economic theory suggests that interest rates are bounded and therefore will return to intermediate levels \cite{executiveofficeofthepresidentoftheunitedstatesLongtermInterestRates2015}.  Since December 2021, the interest rates for long-term UK government bonds have indeed been rising again \cite{smithWhyAreUK22}. Hence, not only the value of the interest rate is uncertain but if and how fast it might return to earlier levels. A scenario analysis can be used to explore the changes in energy system designs under different interest rate evolutions.\newline It has to be underlined that, even though a consistent methodology is applied, the quality of the uncertainty characterisation remains limited by the availability and quality of data. Additional references for demand or price forecasts based on some external models had to be discarded because of methodological changes in the same report series over different years. Furthermore, the assumptions underlying the forecasts are rarely fully disclosed in the reviewed reports. In general, the number of references is limited and therefore does not allow for a comparison of different ranges. None of the references allowed to truly estimate the long-term uncertainty over 15-25 year intervals because the data points were at most 10 years apart. 

\subsection{Comparing the impact of spatial detail to conventional input parameters}\label{Sect:ResultsGSA}
\begin{figure}[H]
\centering
  \includegraphics[width=0.7\textwidth]{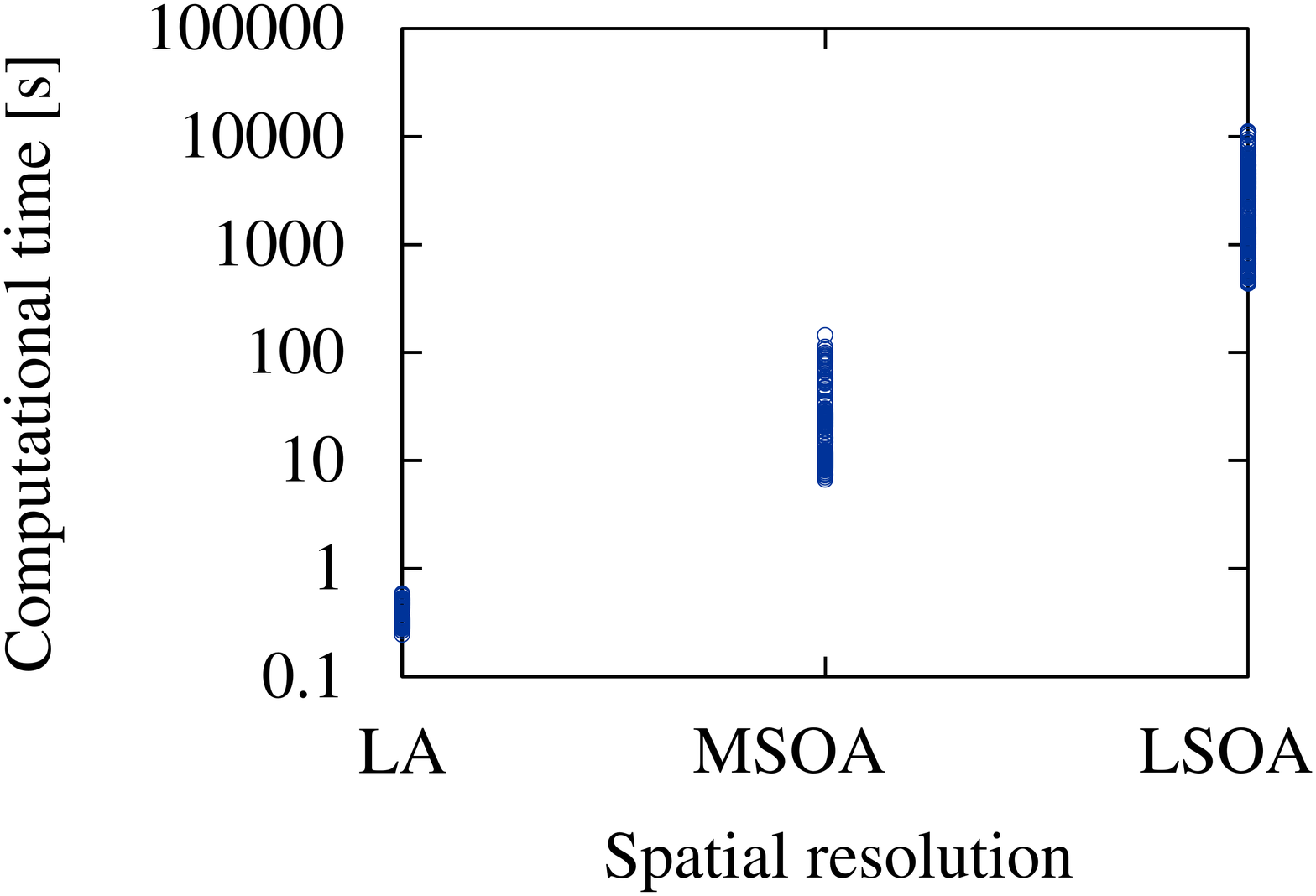}
  \caption{For all sample points in the GSA, the computational time is plotted against the spatial resolution level (increasing from left to right on the x-axis).  Each stepwise increase in spatial resolution increases the computational time by up to two orders of magnitude. At each level of spatial resolution, the variations in computational time stem from the different input parameter vectors.}
  \label{fgr:CompTime}
\end{figure}

The results were generated using $T=24$ trajectories with $p_{\theta} = 4$ for the input parameters and $p_{\varphi} = 3$ for the spatial resolution. $T$ being a multiple of $p_{\theta}$ and $p_{\varphi}$ allowed for uniform sampling of all input factors. In general, $T$ is typically chosen between 10-30 without a formal analysis of the most appropriate number \cite{chitaleEffectiveParameterScreening2017}. For $T(I+1)$ sample points with $I=35$ uncertain input factors, 864 model iterations were needed for the full GSA.\newline In Fig. \ref{fgr:CompTime} the computational time for all sample points is shown as a function of the level of spatial resolution. Each stepwise increase in spatial resolution increases the computational time by up to two orders of magnitude: while the single-node representation (LA) solves within 1 second, the model takes up to 3 hours at the finest resolution (LSOA). Different input parameter vectors cause a variation of the computational time even if the level of spatial resolution is kept constant because the shape of the solution space is altered. It is therefore not possible to determine the total computational time for all sample points \textit{a priori}, but its order of magnitude can be estimated based on the deterministic model formulations. Importantly, the total computational time of the GSA can significantly be reduced by solving the optimisation problem for each sample point in parallel rather than sequentially.
 
\begin{figure*}
\centering
  \includegraphics[width=17.1cm]{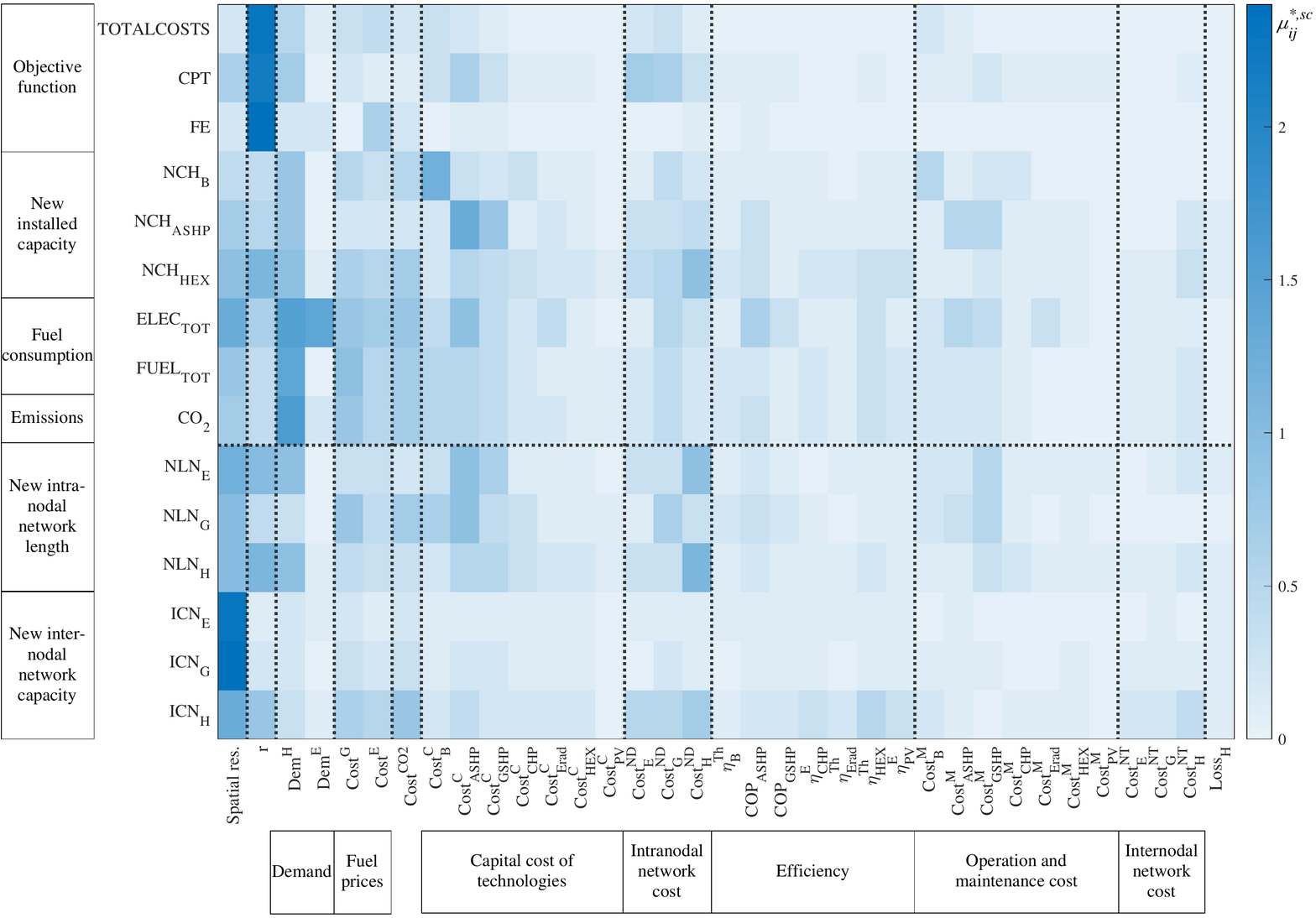}
  \caption{The heat map visualises $\mu_{ij}^{*,sc}$ for a selected subset of design and operation output variables (\textit{y-axis}). The shading directly correlates with the impact of the input factors (\textit{x-axis}). The categories of input factors are ordered from left to right according to their total impact across all of the selected output variables. Spatial resolution ($Spatial~res.$) is mainly important for network capacities, here grouped below the horizontal dotted line. The modelling choice has the highest impact across the output variables together with the conventional input parameters discount rate ($r$) and heat demand ($Dem^H$).}
  \label{fgr:HeatMap}
\end{figure*}

The heat map in Fig. \ref{fgr:HeatMap} allows to compare the impact of all input factors (\textit{horizontal axis}) for a representative sample of the outputs (\textit{vertical axis}). As we use a multi-year model, outputs such as fuel consumption, emissions, new network and technology capacities are summed across the time horizon. The impact is quantified by $\mu_{ij}^{*,sc}$ as introduced in Eq. \ref{eq:scaling}.

\begin{figure}[H]
\centering
  \includegraphics[width=0.8\textwidth]{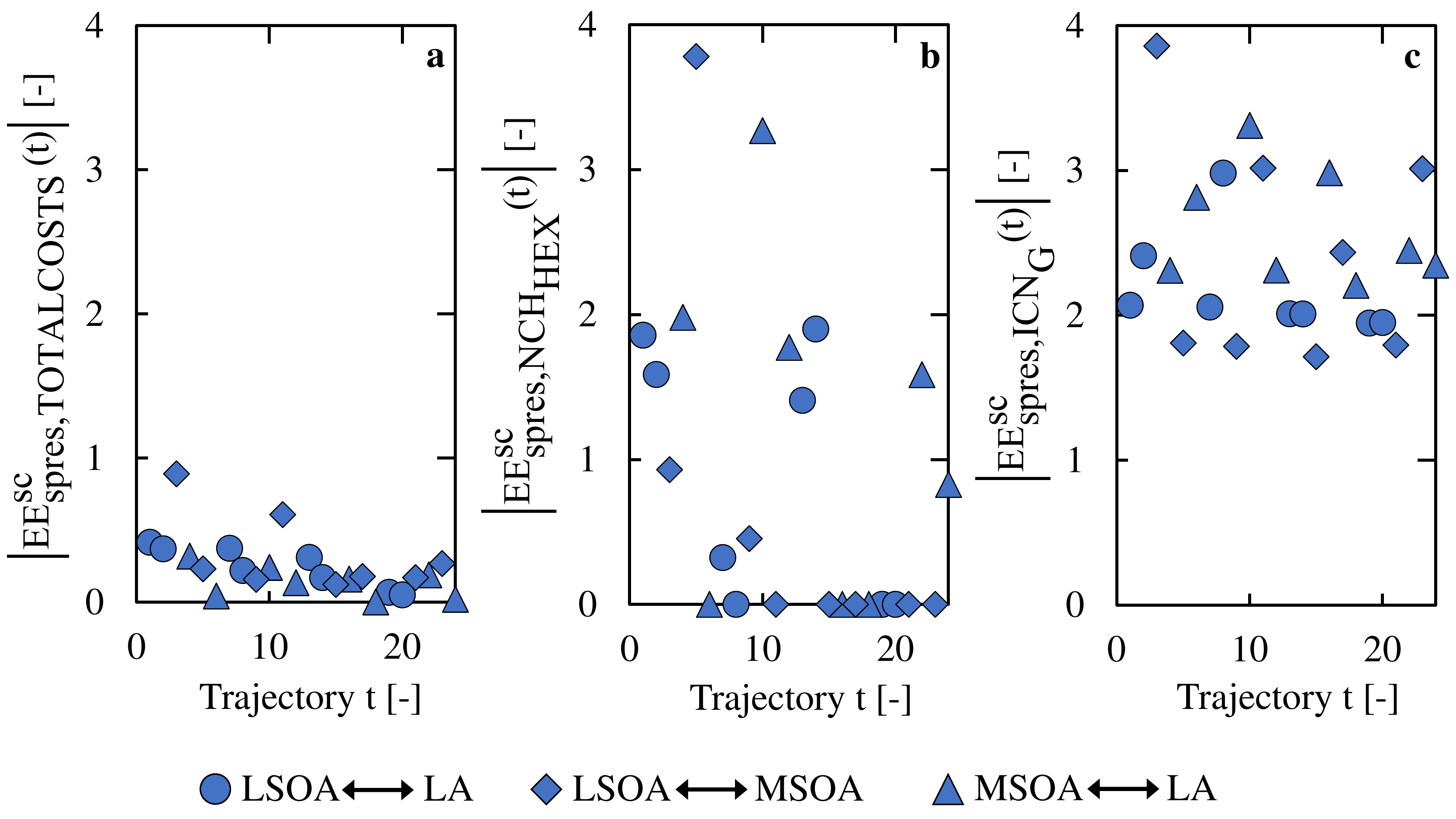}
  \caption{For each trajectory $t$, the scaled absolute EEs for spatial resolution ($\emph{spres}$) with respect to the (a) total system cost ($\emph{TOTALCOSTS}$), (b) heat exchanger capacity ($\emph{NCH}_{\emph{HEX}}$) and (c) internodal natural gas network capacity ($\emph{NCN}_G$) are shown. Each symbol ($\emph{circle}$, $\emph{diamond}$, $\emph{triangle}$) indicates a different transition between two levels of spatial resolution. The impact of $\emph{spres}$ is independent of the input parameters if similar $\left| EE_{\emph{spres},j}(t) \right|$ are obtained for all 24 trajectories.}
  \label{fgr:Comparison}
\end{figure}

\begin{figure}[H]
\centering
  \includegraphics[width=0.8\textwidth]{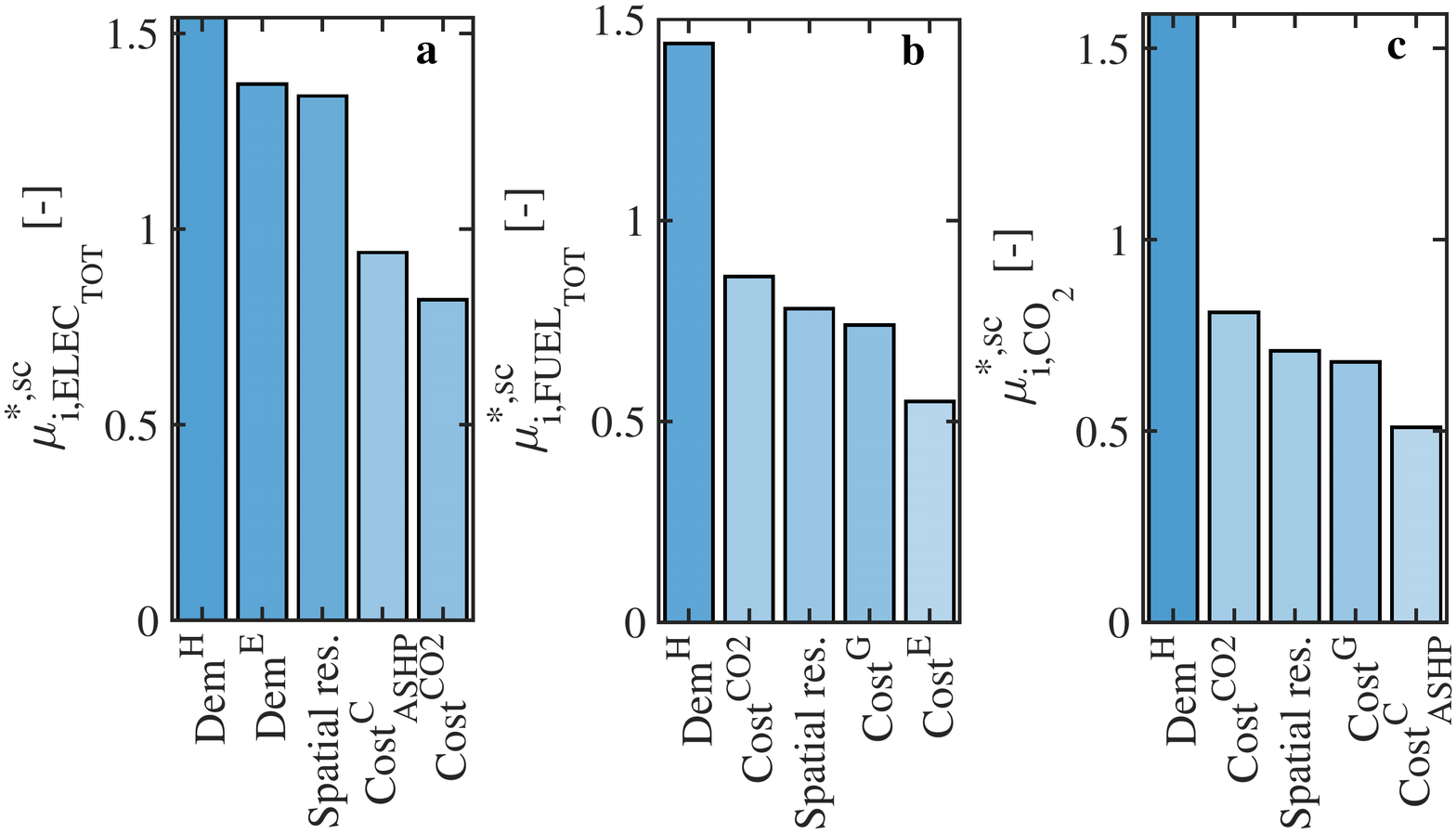}
  \caption{The five highest ranking input factors with respect to the total electricity consumption ($ELEC_{TOT}$), natural gas consumption ($FUEL_{TOT}$) and GHG emissions ($CO_2$) with their explicit $\mu_{ij}^{*,sc}$ are compared. It represents a section of heat map in Fig. \ref{fgr:HeatMap}. The optimal value is dominated by the heat demand. Spatial resolution ranks third across all three outputs.}
  \label{fgr:Barplot}
\end{figure}

For each output variable, the most important input factor is identified by the darkest shade within a row. Given that $\mu_{ij}^{*,sc}$ is dimensionless due to the scaling by $\Tilde{\sigma}_j$ (see Eq. \ref{eq:scaling}), the impact of an input factor can also be compared across different output variables, although in a qualitative way. The darker the column overall, the more important is the input factor across the design and operation variables.\newline Comparing the shade for the different categories of input factors in Fig. \ref{fgr:HeatMap} reveals that only a limited number of input factors have a significant impact. Also in \citet{MAVROMATIDIS2018219} and \citet{MORET2017597} only a subset of the input parameters considered in the GSA were significantly changing the total system cost. The set of negligible input parameters is thereby model dependent. In Fig. \ref{fgr:HeatMap}, the impacts of efficiency, operation \& maintenance cost, internodal network cost and network losses are overall negligible in the HIT model. As summarised in Table \ref{tbl:UncertaintyRanges_UK} and \ref{tbl:UncertaintyRanges_CH}, the magnitude of the uncertainty ranges vary significantly across these input factors. As input factors with large uncertainty ranges but negligible impact are found, this shows that the impact of an input factor is not necessarily correlated to the size of its uncertainty range.
The observation suggests that the savings in operational costs due to increased efficiencies ($\eta^{\emph{Th}}_\emph{tech}$ /$\emph{COP}_\emph{tech}$/$\eta^E_\emph{tech}$) or reduced network losses ($\emph{Loss}_H$) cannot outweigh the differences in capital cost between the technologies in the HIT model. The annual operation \& maintenance cost ($\emph{Cost}^{M}_{\emph{tech}}$) are implemented as 5-10\% of the capital cost and therefore represent only a minor contribution to the total system cost. The negligible impact of internodal network costs ($Cost^{NT}_{E/G/H}$) might be a consequence of the smaller absolute uncertainty range compared to the intranodal network costs ($Cost^{ND}_{E/G/H}$).\newline Summing the $\mu_{ij}^{*,sc}$ within a column for all output variables presented in Fig. \ref{fgr:HeatMap}, spatial resolution turns out to be the most relevant input factor, followed by the discount rate and the annual heat demand. While the spatial resolution and the discount rate both can reach $\mu_{ij}^{*,sc}~>~2$, their impacts strongly depend on the output variable.\newline The discount rate dominates the total system cost ($\emph{TOTALCOSTS}$), capital cost ($\emph{CPT}$) and fuel cost ($\emph{FE}$) because the objective function and its cost terms are calculated as net present values (NPV). The NPV is determined as the sum of the discounted costs in each investment period across the time horizon and was previously observed to be highly sensitive to discount rate assumptions \cite{GARCIAGUSANO201656}. In fact, higher discount rates favour capital intensive investments in the future \cite{L_ffler_2021} but in return penalise future revenues and cost savings that are achieved through early investments in more efficient and less carbon-intensive technologies. The impact of the discount rate on time value of money over time makes it also the most important parameter for the optimal heat network deployment.  The greatest cost savings are achieved by early investments into the heat pipeline network given its assumed lifetime of 50 years. Hence, heat networks are favoured at low discount rates whereas high discount rates favour late investments into heat pumps.\newline The ranking with respect to the newly installed capacities of natural gas boilers ($\emph{NCH}_B$) and air-sourced heat pumps ($\emph{NCH}_\emph{ASHP}$) in Fig. \ref{fgr:HeatMap} shows that the capital cost of the technology itself is the most important input factor. $\emph{NCH}_B$ is observed to be insensitive to the carbon tax ($\emph{Cost}^{\emph{CO2}}$) suggesting that the price increase across the time horizon is too small to cause a significant shift away from individual gas boilers towards more efficient DH boilers or electrification via heat pumps. In contrast to the individual heating technologies, the capacity of heat exchangers ($\emph{NCH}_{HEX}$) is driven by the intranodal pipeline cost of heat networks ($\emph{Cost}^{ND}_H$) rather than the capital cost of the heat exchangers ($\emph{Cost}^{C}_{HEX}$) themselves. The observation suggests that the high upfront of the heat networks limits the deployment of DH systems.\newline As Fig. \ref{fgr:HeatMap} highlights, the spatial resolution has negligible impact on the total system cost, but is by far the most important input factor for the installed capacity of internodal networks ($\emph{ICN}_{E/G/H}$). Out of all heat technologies, the installed capacity of heat exchangers ($\emph{NCH}_\emph{HEX}$) is most sensitive to the choice of spatial resolution. For $\emph{NCH}_\emph{HEX}$, spatial resolution ranks third after the discount rate and pipeline cost. As discussed in Section \ref{Sect:LitReview}, previous publications on the impact of spatial resolution in power systems \cite{FRYSZTACKI2021116726, SIMOES2017183,7741996} made similar observations, reporting minor changes in the total system cost but greater changes to the capacities of technologies and networks. This qualitative trend for the impact of spatial resolution across different types of output variables is further supported by Fig. \ref{fgr:Comparison} as the impact is analysed for different input vectors $\boldsymbol{\mathrm{\theta}}$. For each of the 24 trajectories, the scaled absolute EEs ($\left| EE^{sc}_{ij}(t)\right|$) of the total system cost (\emph{TOTALCOSTS}), heat exchanger capacity ($\emph{NCH}_\emph{HEX}$) and internodal gas network capacity ($\emph{ICN}_G$) for changes in the spatial resolution by $\Delta_{\varphi}$ are shown. Different symbols are used to differentiate the transitions between two levels of spatial resolution. In Fig. \ref{fgr:Comparison}a, small absolute EEs are observed for every symbol and all trajectories. Hence, the spatial resolution remains negligible for the total system cost, independent of the change in spatial resolution or the input vector. Similarly, spatial resolution is always important for the sizing of the internodal gas network capacity ($\emph{ICN}_G$) as shown in Fig. \ref{fgr:Comparison}c. In contrast, $\left| EE^{sc}_{\emph{spres},\emph{NCH}_\emph{HEX}}(t)\right|=0$ and $0 < \left| EE^{sc}_{\emph{spres},\emph{NCH}_\emph{HEX}}(t)\right|$ both can be observed in Fig. \ref{fgr:Comparison}b. For each type of transition between two levels of spatial resolution, the scaled absolute EE varies dependent on the trajectory. This shows that the impact of spatial resolution on the installed capacity of heat exchangers ($\emph{NCH}_\emph{HEX}$) is strongly dependent on the value of the input parameters. 
As observed in Fig. \ref{fgr:HeatMap}, spatial resolution introduces the greatest changes to the internodal electricity, gas and heat network capacities ($ICN_{E/G/H}$). By definition, the internodal network capacity is zero for single-node representations. In the HIT model, internodal networks are used to distribute gas and electricity to each node within the area as only the nodes in the North and South are directly connected to the transmission grids. Therefore, internodal network capacities ($ICN_{E/G/H}$) are generally expected to increase when increasing the resolution to MSOA or LSOA level. However, as exemplified for $ICN_{G}$ in Fig. \ref{fgr:Comparison} , the EEs for changes between the finest and medium level of spatial resolution (\emph{diamond}) are comparable to the EEs for transitions involving the single-node level (\emph{triangle, circle}). The large impact of spatial resolution on $ICN_{E/G/H}$ across all transitions is explained by the averaging of the electricity and heat demands across LSOA that are aggregated to MSOA or the LA (see Fig. \ref{fgr:Winchester}). Network capacities are generally sized to meet the peak demand. The averaging of LSOA demands reduces the peak demands and therefore results in an undersizing of the internodal network capacities $ICN_{E/G/H}$ in aggregated system representations. Hence, the highest spatial resolution using LSOA representation is required to accurately size the network capacities in the HIT model.\newline Spatial resolution remains the most important input factor for the intranodal network length of the electricity and gas networks ($\emph{NLN}_{E/G}$). By increasing the spatial resolution from MSOA to LSOA, the deployment of DH systems is focused on the small LSOA with high heat demand densities. In return, gas and electricity networks are deployed in larger LSOA with lower heat demand densities. As the intranodal network length $\emph{NLN}_{E/G}$ is proportional to the total road length within the node, increased spatial resolution leads to increased $\emph{NLN}_{E/G}$. For the intranodal heat network length $\emph{NLN}_{H}$, the impact of the pipeline investment cost ($\emph{Cost}^{ND}_H$) and the discount rate ($\emph{r}$) however outweigh the impact of spatial resolution. Nevertheless, as all three  types of intranodal networks are sensitive to the spatial resolution, it is recommended to determine their design at the highest spatial resolution.\newline A more detailed representation of the input factor ranking for the total electricity consumption ($ELEC_{TOT}$), natural gas consumption ($FUEL_{TOT}$) and GHG emissions ($CO_2$) shown in Fig. \ref{fgr:HeatMap} is provided by Fig. \ref{fgr:Barplot}. 
The heat demand has the highest impact with respect to these three operational variables. Higher heat demands lead to higher natural gas consumption ($FUELTOT$) that is associated with higher CO$_2$ emissions ($CO_2$). Even though the heat demand also ranks first for the electricity consumption, it shows smaller differences to the second and third rank because the heat demand changes are attenuated by the high COP of heat pumps. For the cost assumptions used in the HIT model, solar PV is not cost-competitive. Without the decentralised generation of solar power, an increase in the annual electricity demand directly translates into an increased total electricity consumption from the grid, causing a sensitivity despite the comparably low uncertainty ranges.\newline In a cost optimisation, higher prices of an energy vector would be expected to shift the consumption to alternative energy sources. In Fig. \ref{fgr:Barplot}, spatial resolution is however shown to rank even higher than the fuel prices across all three output variables. On average, higher spatial resolutions decrease $\emph{ELEC}_{TOT}$ and increase $\emph{FUELTOT}$ with $\emph{CO}_2$ due to the increased deployment of DH systems fuelled by large-scale natural gas boilers.\newline For $FUELTOT$ and $CO_2$, $\mu_{ij}^{*,sc}$ of the carbon price, spatial resolution and natural gas prices are very similar. The ranking is therefore expected to be model dependent and should be confirmed for other uncertainty bounds as the ranking of an input parameter $\theta_k$ has previously been shown to vary for different $R_k^{\%,\emph{min}}$ and $R_k^{\%,\emph{max}}$ \cite{MORET2017597}.

\subsection{Balancing detail and uncertainty depending on the research question}
The results in Section \ref{Sect:ResultsGSA} show that the impact of a given modelling choice -- in our case, spatial resolution -- is dependent on the type of output variable of interest. Therefore, the decision on whether to invest the limited computational resources on increasing resolution or on optimisation under uncertainty depends on the research question. In the following, guidance for model application and development is derived based on the ranking of input factors for the HIT model used in this paper.\newline As spatial resolution had negligible impact on the total system cost, a single node representation would already be suitable to compare the total system cost for different LAs or to determine the optimal total system cost for uncertain interest rates using robust optimisation. As the impact has also resulted to be negligible in previous publications targeting spatial resolution, the adequacy of single node representations for total system cost calculations might also apply to ESM other than the HIT model.\newline In Fig. \ref{fgr:Comparison}b, the total heat exchanger capacity determined at MSOA level tends to be the same at LSOA level  as $\left| EE_{\emph{spres},\emph{NCH}_{\emph{HEX}}}(t) \right| = 0$ for most transitions that occur between MSOA and LSOA (\emph{diamond}). Furthermore, if $0 < \left| EE_{\emph{spres},\emph{NCH}_{\emph{HEX}}}(t) \right|$ is observed, the $\left| EE_{\emph{spres},\emph{NCH}_{\emph{HEX}}}(t) \right|$ values are generally smaller than for transitions between MSOA and LA. An outlier (\emph{diamond}, 5$^{\text{th}}$ trajectory) is identified when district heating only becomes cost-competitive at LSOA level because high discount rate make heat electrification more economical. Otherwise, these observations suggest that introducing some spatial resolution by moving from a single node (LA) to a 10-node (MSOA) representation is sufficient to determine the split in the capacities of individual and district heating technologies. Keeping the system representation at MSOA level allows saving computational time for a robust optimisation for uncertain capital costs and discount rates. Other analyses on the capacity of heating technologies such as the carbon prices required to motivate households to replace their natural gas boilers or the necessary learning rates for heat pumps to become cost-competitive without subsidies could similarly be answered using an intermediate spatial resolution.\newline As discussed based on Fig. \ref{fgr:Barplot}, the energy consumption and GHG emissions show similar sensitivities to spatial resolution, energy and carbon prices. Therefore, an accurate estimate would require increasing resolution as well as including uncertainty. However, as the focus of the HIT model lies on the capacity sizing, the temporal representation is simplified. While the level of temporal detail is sufficient for an estimate of the ratio between capital and operating costs for each technology, it is insufficient to provide accurate estimates of the energy consumption and GHG emissions.\newline Spatial resolution is irreplaceable for a correct sizing of the internodal network capacities. Investing the computational resources for spatially explicit energy systems optimisation is therefore especially relevant for systems in which technology deployment requires significant network reinforcement or expansion. In the case study, the LSOA nodes were aggregated according to administrative boundaries to MSOA and LA level, respectively. At lower levels of detail (LA and MSOA), the averaging of the demand profiles across LSOA with very different heat demand densities causes the internodal network to be undersized. Instead, clustering the nodes according to heat demand densities and spatial proximity could better approximate the peak heat demand and save computational time. Using visual inspection to cluster the adjacent LSOA within the same linear heat density range in Fig. \ref{fgr:Winchester}, the heat demand distribution could be captured with 17 instead of 49 nodes. In the HIT model, the investment decisions in the end-use technologies made at MSOA level were insensitive to a further increase in spatial resolution. Therefore, it would be possible to separate the problem into the design of supply technologies at MSOA level followed by a refined network design at LSOA level using fixed technology capacities.

\section{Conclusion}
The level of detail and uncertainty of input parameters were both observed to affect the type and size of the technology investments in long-term ESM including high shares of renewable energy. Computational tractability however limits the optimisation of highly detailed models under uncertainty. Although separate methodologies assessing the impact of detail or uncertainty have been presented in the literature, no methodological framework allowed for the comparison of both.\newline The novel application of GSA presented in this work closes this gap by considering modelling choices like the level of detail alongside conventional input parameters. A discrete uniform distribution is assigned to a set of model formulations differing in a single modelling choice, with each model formulation having equal probability. By applying global sensitivity analysis, the impact of the modelling choice against the uncertainty in input parameters can be compared for each output variable of interest.\newline The method was illustrated by comparing the impact of spatial detail and uncertainty in an urban heat supply model for the UK. In the uncertainty characterisation step, larger ranges for the market-driven investment cost of technologies and fuel prices than for the efficiencies of technologies were found. The case study underlined that the relevance of spatial detail or uncertainty is dependent on the output variable of interest. Spatial resolution was identified to be most relevant for network capacities, while its impact on the total system cost was negligible. The investment costs of the individual heat technology were most relevant for the capacity sizing. But given the high capital costs of district heating networks, their capacity was most sensitive to the uncertainty in the discount rate. Overall, moving from no spatial resolution (1-node model) to medium spatial resolution (10-node) is shown to have a significant impact on determining the optimal investment strategy. Further refinement to a high level of spatial resolution (49 nodes) only seems justifiable for the detailed design of networks as high computational cost are incurred for otherwise marginal changes to the investment strategy.\newline Our findings are aligned with the existing literature. Comparing the ranking for different output variables indicates that neither a highly detailed deterministic nor a coarse optimisation under uncertainty will allow to determine all output variables accurately. Therefore, the first critical step for model development remains formulating a research question and identifying the key output variable(s). In a second step, applying our method can make it possible to strike the necessary balance between detail and uncertainty in the ESM development.

\section*{Acknowledgements}
This research was funded by the UK Engineering and Physical Sciences Research Council (EPSRC) [grant number EP/R045518/1]. The authors would also like to thank Yogesh Khare (The Everglades Foundation) and Rafael Munoz-Carpe\~na (University of Florida) for answering our questions regarding the Elementary Effects packages. Francisca Jalil-Vega (University of Bristol) and Adam D. Hawkes (Imperial College London), thank you for providing the HIT model and the geographic information system files for Winchester.

\section*{Author Contributions}
Maria Yliruka: Conceptualization, Methodology, Data curation, Formal analysis, Validation, Software, Writing-Original draft preparation, Visualization, Investigation. Stefano Moret: Supervision, Conceptualization, Validation, Writing-Reviewing and Editing. Nilay Shah: Supervision, Project administration, Funding acquisition, Writing-Reviewing and Editing.

\section*{Conflicts of interest}
There are no conflicts to declare.



\pagebreak

\appendix

\section{Uncertainty characterisation for the UK energy system}\label{Sct:AppendixA}
\subsection{Final energy demand}\label{Sct:AppDemand}
\begin{figure}[H]
\centering
\includegraphics[width=0.9\textwidth]{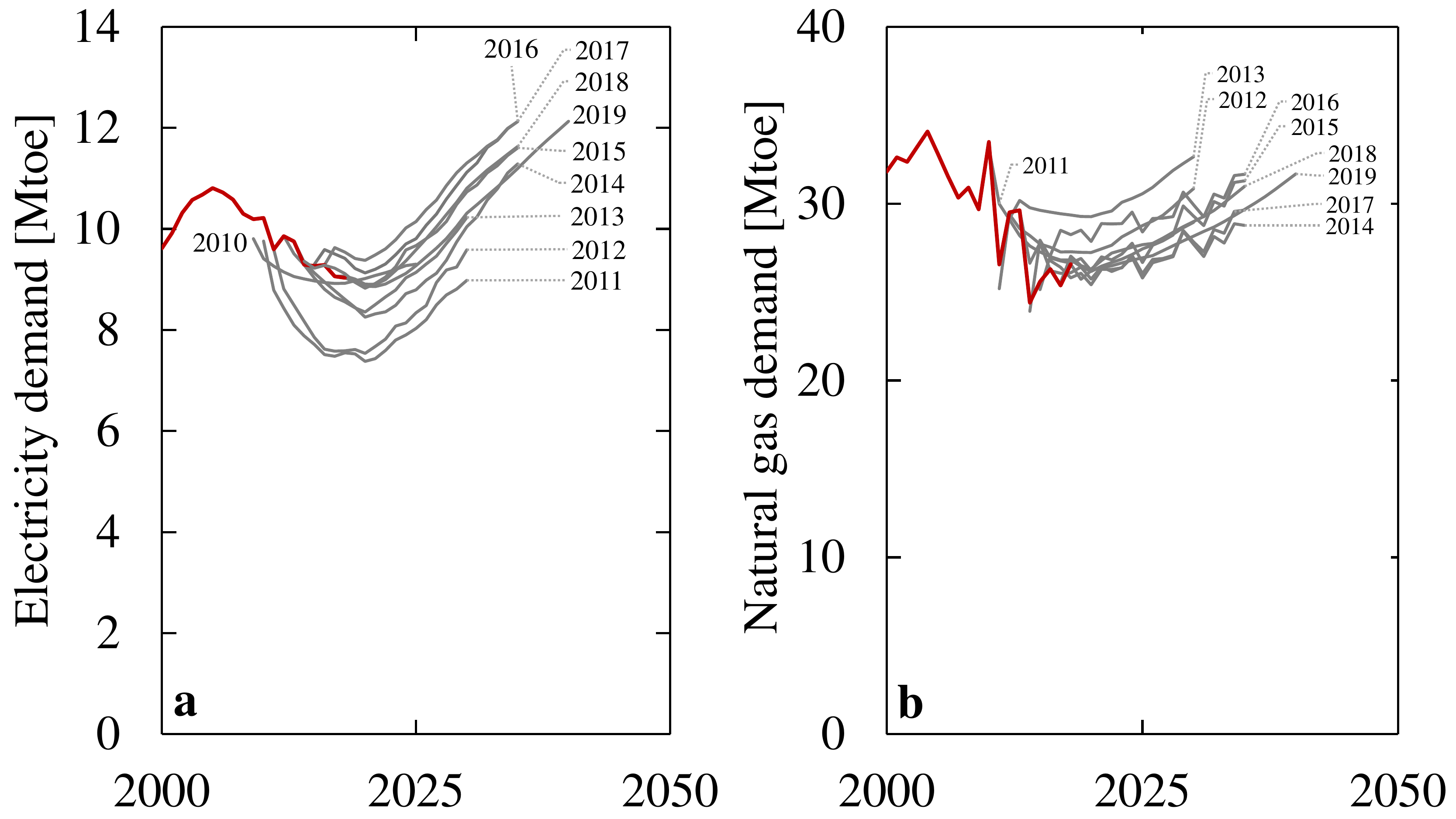}
\caption{The domestic electricity (a) and gas (b) demand projections (\emph{grey lines}) are contrasted to the actual values (\emph{red line}) as published in the UEP reports 2010-2020 \cite{departmentforbusinessenergyindustrialstrategyCollectionEnergyEmission2010a}. For each projection, the corresponding year of publication is stated.}
\label{fgr:DomesticDemand}
\end{figure}

The uncertainty ranges for the final energy demands $Dem^E$ and $Dem^H$ are based on the annual `Updated energy and emission projection' (UEP) reports between 2010-2020 published by the Department for Business, Energy and Industrial Strategy (BEIS) \cite{departmentforbusinessenergyindustrialstrategyCollectionEnergyEmission2010a}. The individual reports comprise actual and forecast values for the annual final energy demand in different sectors including the domestic, commercial and public sector split according to the type of fuel.\newline For the HIT model, the uncertainty ranges of the domestic electricity and natural gas demands were of interest. Their projections are contrasted to the actual demand evolution in Fig. \ref{fgr:DomesticDemand}. For both final energy demands, an overall reduction since 2000 is observed. The uneven profile of the actual natural gas demands is caused by its weather dependency given that it is mostly used for space heating.
The final electricity and natural gas demands of the domestic sector in 2008 were consistently reported within all the UEP reports and were therefore used as criteria to verify the continuity between the reports. As the demand values for 2008 stated in the UEP reports of 2010 and 2011 differed from the UEP reports of 2012-2020, the two reports were excluded from the analysis.\newline 
To calculate the relative deviation, $\Delta_{\%}$, tabulated in Table \ref{tbl:FinalElecDemand} and \ref{tbl:FinalNGDemand}, the forecast and actual demands in the `reference scenario'\footnote{The names of the scenarios are inconsistent. For 2010 `Central prices' and for 2011 and 2012 `Central scenario' are used.} were compared for UEP reports that were 5 years apart. For example, the forecast for 2016 in the UEP report of 2012 was compared to the actual demand in 2016 published in the UEP report of 2017. Note, that the actual demands are reported with hindsight. Therefore, the 2017 UEP report states the actual value for 2016 and makes projections for 2017-2035. \newline For a conservative estimate, the uncertainty ranges were based on the extreme values $\Delta_{\%}$ in Table \ref{tbl:FinalElecDemand} and \ref{tbl:FinalNGDemand}, respectively. However,  the forecast error significantly improved over time for the domestic electricity demand. It was assumed that this improvement was a result of methodological changes and therefore would be systematic. Therefore, $\Delta_{\%}$ for 2016 was excluded. In a second step, the extreme values for $\Delta_{\%}$ derived for a 5-year interval were linearly extrapolated to a 15-year time horizon, as \cite{MORET2017597} had previously observed increases forecast errors for final energy demands over time. Hence, the final uncertainty ranges of $R^{\%,\emph{min}}_{\emph{Dem}^E}=0\%$ and $R^{\%,\emph{max}}_{\emph{Dem}^E}=15\%$ for domestic electricity demands and $R^{\%,\emph{min}}_{\emph{Dem}^H}=-40\%$ and $R^{\%,\emph{max}}_{\emph{Dem}^H}=6\%$ for domestic natural gas demands are found.
\begin{table}[H]
\caption{Comparison of the forecast and actual final electricity demands in the domestic sector for 2016-2019. The domestic electricity demands are consistently underestimated in the projections published annually by BEIS \cite{departmentforbusinessenergyindustrialstrategyCollectionEnergyEmission2010a}}
\label{tbl:FinalElecDemand}
\centering
\begin{tabular}{c|c|c|c}
Year & Actual [ktoe] & Forecast [ktoe] & $\Delta_{\%}$ [$\%$] \\
\hline
2016 & 9283.86 & 7619.79 & +22 \\
2017 & 9062.41 & 8645.93 & +5 \\
2018 & 9033.97 & 8608.80 & +5 \\
2019 & 9033.92 & 8944.07 & +1 \\
\end{tabular}
\end{table}
\begin{table}[H]
\caption{Comparison of the forecast and actual final natural gas demands in the domestic sector for 2016-2019 published in the annual `Updated energy and emission projection' by BEIS \cite{departmentforbusinessenergyindustrialstrategyCollectionEnergyEmission2010a}}
\label{tbl:FinalNGDemand}
\centering
\begin{tabular}{c|c|c|c}
Year & Actual [ktoe] & Forecast [ktoe] & $\Delta_{\%}$ [$\%$] \\
\hline
2016 & 26773.44 & 27534.61 & -3 \\
2017 & 25540.43	& 29427.06 & -13 \\
2018 & 26584.00 & 26094.86 & +2 \\
2019 & 26583.86 & 26901.36 & -1 \\
\end{tabular}
\end{table}
The uncertainty range for the domestic natural gas demand is larger than for the domestic electricity demand as expected given that the majority of natural gas is consumed for weather-dependent space heating, which in the UK is mostly provided by natural gas boilers.
\newline Overall, the documentation of BEIS does not provide many details on the assumptions underlying the projections shown in Fig. \ref{fgr:DomesticDemand}. Economic growth, fuel prices and ambient temperatures are stated to be key drivers \cite{departmentforbusinessenergyindustrialstrategyUpdatedEnergyEmissions2017}. The demand increase past 2026 is explained by slow diminishing of the effects of policies and macroeconomic drivers \cite{departmentforbusinessenergyindustrialstrategyUpdatedEnergyEmissions2020}. From the documentation it remains unclear whether the decarbonisation of the domestic heating demands was taken into account given that a the natural gas demand is projected to increase beyond 2025. Similarly, it is unclear whether the increase in the final electricity demand is also based on assumptions regarding the electrification of domestic transport or heat in the projections shown in Fig. \ref{fgr:DomesticDemand}a. Without details on the underlying assumptions, the users cannot decide whether the demand projections and their uncertainty ranges are aligned with the assumptions applied in their work.\newline Developing final energy demand projections requires access to the historical and actual national energy demands split by sector and by region. Therefore, only energy suppliers and governmental bodies are in a position to publish such analysis. While the Future Energy Scenarios (FES) of National Grid are often used as a reference for demand projections, the reports between 2011-2019 \cite{nationalgridFutureEnergyScenarios2011} could not be used to perform a similar uncertainty characterisation because the reporting was inconsistent. Therefore, the UEP reports remain the only reference despite the short-comings in their documentation.

\subsection{Carbon tax}\label{Sct:AppCO2}
\begin{figure}[H]
\centering
\includegraphics[width=0.9\textwidth]{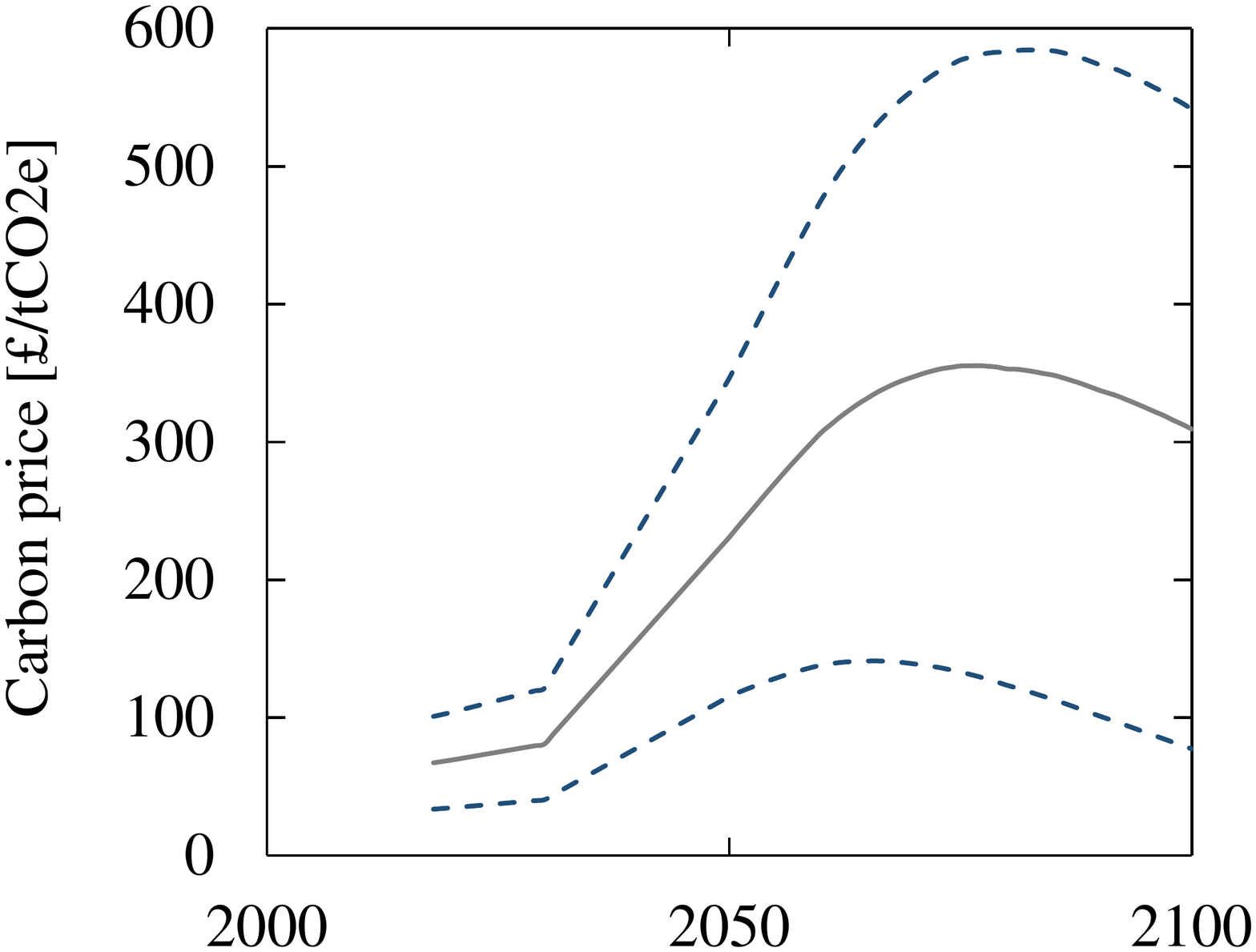}
\caption{The non-traded carbon price and its sensitivity bounds as suggested by BEIS \cite{departmentforbusinessenergyindustrialstrategyGreenBookSupplementary2012} are shown. A linear increase in the carbon price at different rates for 2018-2030 and 2030-2050 are observed. Until 2050, the lower and upper bound of the sensitivity range (\emph{dashed lines}) correspond to 50\% and 150\% of the central carbon price (\emph{full line}), respectively.}
\label{fgr:CarbonTax}
\end{figure}

In the HIT model \cite{JALILVEGA20181051}, the central projection of the non-traded carbon price in the supplementary guidance to the Green Book published by BEIS \cite{departmentforbusinessenergyindustrialstrategyGreenBookSupplementary2012} was used as input value for the carbon tax between 2020-2050. Therefore, the uncertainty range was also determined based on the sensitivities for the non-traded carbon price suggested by BEIS. The evolution of the central price and its sensitivity bounds are shown in Fig. \ref{fgr:CarbonTax}. \newline Aligned with the time horizon considered in the HIT model, the analysis was limited to the projections until 2050. As Fig. \ref{fgr:CarbonTax} shows, the time horizon until 2050 can be divided into two intervals, 2018-2030 and 2030-2050, in which the carbon price as well as its sensitivity bounds increase linearly. For each year between 2018 and 2050, the lower and upper sensitivity bound correspond to 50\% and 150\% of the central carbon price, respectively. Hence, a symmetric uncertainty range of $R^{\%,\emph{min}}_{\emph{Cost}^{\emph{CO}_2}}=-50\%$ and $R^{\%,\emph{max}}_{\emph{Cost}^{\emph{CO}_2}}=50\%$ was determined.

\subsection{Fuel prices}\label{Sct:AppFuelPrices}
\begin{figure}[H]
\centering
\includegraphics[width=0.9\textwidth]{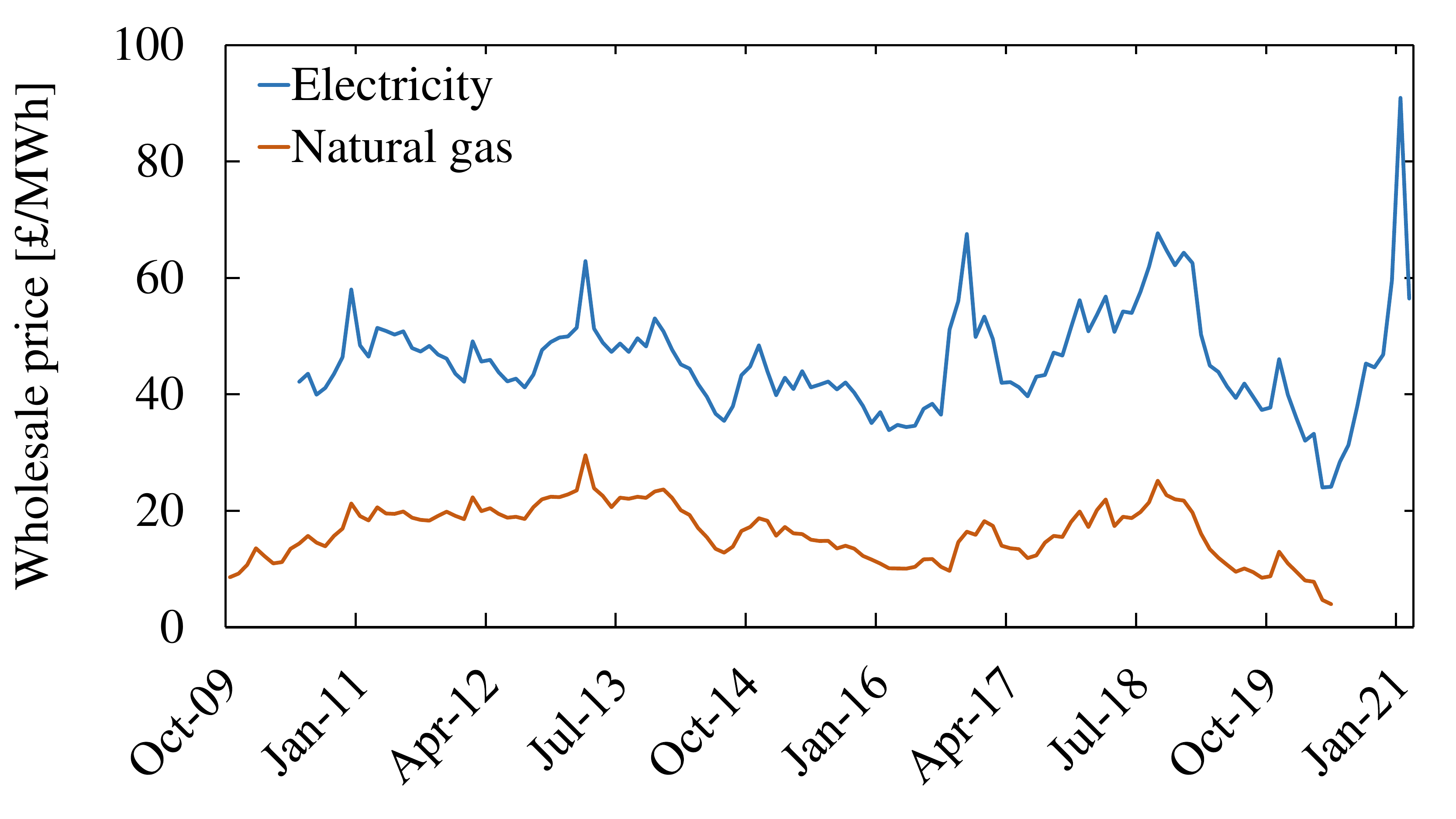}
\caption{The monthly average electricity and natural gas wholesale prices in the UK from October 2009 until February 2021 as published by Ofgem \cite{officeofgasandelectricitymarketsWholesaleMarketIndicators2009} are shown. The correlation between both prices can be explained by the increased use of natural gas for baseload electricity demands in the UK \cite{gisseyWholesaleCostReflectivity2018}.}
\label{fgr:FuelPrices}
\end{figure}

The domestic end-users represented in the HIT model pay retail prices that include costs such as network, operating, environmental/social costs and VAT additional to the wholesale price. According to the Office of Gas and Electricity Markets (Ofgem), 34\% of the retail electricity price was determined by the wholesale electricity price in 2019 whereas for natural gas, the share increases to 46\% \cite{officeofgasandelectricitymarketsInfographicBillsPrices2021}. The additional costs were assumed to be less volatile than the wholesale prices. The monthly average wholesale electricity and gas prices between 2009 and 2020 published by Ofgem \cite{officeofgasandelectricitymarketsWholesaleMarketIndicators2009} were therefore used to determine the uncertainty ranges for $\emph{Cost}^E$ and $\emph{Cost}^G$. These prices refer to day-ahead contracts and in case of electricity exclude the peak rates between 7 am and 7 pm.\newline First, the monthly average wholesale prices were aggregated into an annual average value. Then, the actual wholesale prices in 2015-2019 were compared across a 5-year time horizon to the prices in 2010-2014, here referred to as forecast values. This comparison is summarised in Table  \ref{tbl:Ofgem_WholeSalePrices_El} and \ref{tbl:Ofgem_WholeSalePrices_NG}. The year stated in these tables always refers to the year of the actual wholesale prices. For each year, the deviation of the actual value from the forecast value is expressed in relative terms as $\Delta_{\%}$. For a conservative estimate of the uncertainty bounds, the extreme values of $\Delta_{\%}$ were used to define $R_{\emph{Cost}^E}^{\%,\emph{min}}=-12\%$ and $R_{\emph{Cost}^E}^{\%,\emph{max}}=15\%$ for the electricity prices (see Table \ref{tbl:Ofgem_WholeSalePrices_El}) as well as $R_{\emph{Cost}^G}^{\%,\emph{min}}=-38\%$ and $R_{\emph{Cost}^G}^{\%,\emph{max}}=0\%$ for the natural gas prices (see Table \ref{tbl:Ofgem_WholeSalePrices_NG}).  
\begin{table}[H]
\caption{Comparison of the actual and forecast annual electricity wholesale prices across a 5-year time horizon. Forecast values tend to equally over- and underestimate the actual wholesale electricity prices.}
\label{tbl:Ofgem_WholeSalePrices_El}
\centering
\begin{tabular}{c|c|c|c}
\toprule
Year & Actual [£/MWh] & Forecast [£/MWh] & $\Delta_{\%}$ [$\%$] \\\hline
2016 & 42.63 & 48.20 & -12 \\
2017 & 46.31 & 45.21 & +2 \\
2018 & 58.23 & 50.80 & +15 \\
2019 & 36.94 & 42.43 & +3 \\
2020 & 41.53 & 40.75 & -9 \\
\bottomrule
\end{tabular}
\end{table}
\begin{table}[H]
\caption{Comparison of the actual and forecast annual natural gas wholesale prices across a 5-year time horizon. The year of the actual natural gas price is stated. As $\Delta_{\%}$ is negative for every year, forecast prices consistently overestimate the actual natural gas prices.}
\label{tbl:Ofgem_WholeSalePrices_NG}
\centering
\begin{tabular}{c|c|c|c}
\toprule
Year & Actual [£/MWh] & Forecast [£/MWh] & $\Delta_{\%}$ [$\%$] \\\hline
2015 & 14.56 & 14.49 & 0 \\
2016 & 11.83 & 19.22 & -38 \\
2017 & 15.38 & 20.38 & -25 \\
2018 & 20.60 & 23.24 & -11 \\
2019 & 11.85 & 16.54 & -31 \\
\bottomrule
\end{tabular}
\end{table}
Fig. \ref{fgr:FuelPrices} shows that the fluctuations in the wholesale electricity price are strongly correlated with the natural gas price fluctuations because natural gas is increasingly used to meet the baseload electricity demands \cite{gisseyWholesaleCostReflectivity2018}. Given that the UK is a net importer of natural gas (64\% in 2018/2019 \cite{officeofgasandelectricitymarketsStateEnergyMarket2019}), both fluctuations can likely be explained by global events. Falling oil prices and weak global demands caused a downward trend between 2013 and 2016 \cite{officeofgasandelectricitymarketsStateEnergyMarket2017} and the price drop in 2020 is explained by the weak global demand during the pandemic. Specifically in the UK, the political uncertainty and the Sterling's depreciation following the EU referendum led to higher fuel prices and more volatility after June 2016 \cite{gisseyWholesaleCostReflectivity2018}. Additionally, the natural gas prices spiked during the cold winters in 2012 and 2013 \cite{officeofgasandelectricitymarketsStateEnergyMarket2015}. Overall, the trend in natural gas prices leads to an overestimation of the actual prices when historical values are used. However, in the future, the trend might not persist as the natural gas demand is expected to grow globally until 2037 \cite{mckinseycompanyGlobalGasOutlook2021} and the increased demands could lead to higher prices if the global production capacity cannot be increased.
\subsection{Interest rate}\label{Sct:AppDiscRate}
\begin{figure}[H]
\centering
\includegraphics[width=0.9\textwidth]{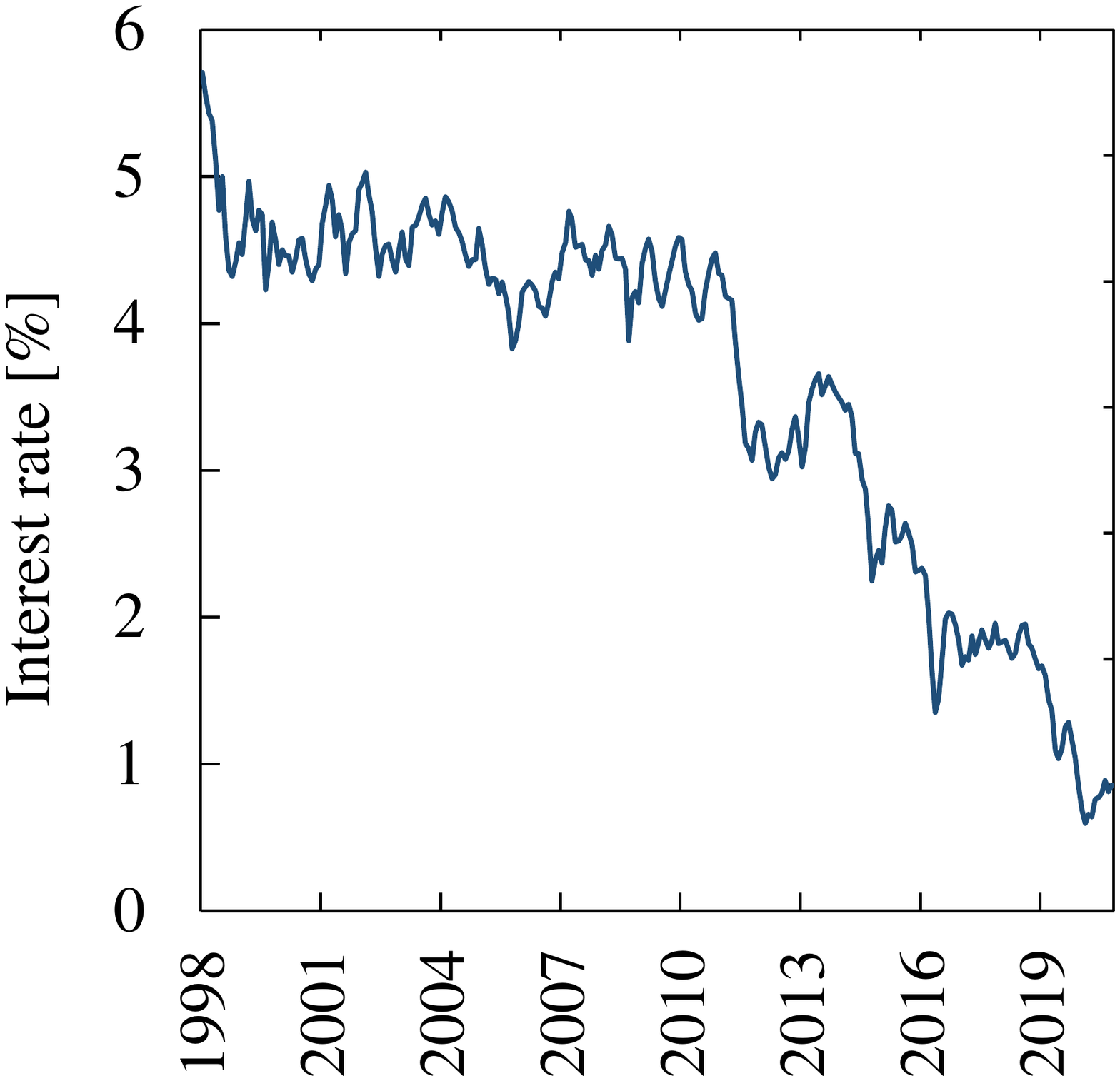}
\caption{The monthly average interest rates for long-term UK government bonds from 1998 until 2020 as published by the DMO \cite{debtmanagementofficeHistoricalAverageDailyn.d.} are shown. Similarly to other OECD countries, the long-term nominal rates steadily declined since 1995 \cite{executiveofficeofthepresidentoftheunitedstatesLongtermInterestRates2015}.}
\label{fgr:InterestRate}
\end{figure}

The HIT model adopts a central planner's perspective to determine the cost-optimal heat decarbonisation strategy. In its application to local authorities across the UK \cite{JALILVEGA2018339}, each local authority can be seen as the central planner. Local authorities can take a loan to finance large infrastructure projects either from private lenders or the Public Works Loan Board (PWLB), a lending facility operated by the UK Debt Management Office (DMO) and financed via the National Loans Fund \cite{debtmanagementofficePWLBLendingFacilityn.d.}. The potential private lenders can be a very heterogeneous group that is difficult to characterise. Therefore, only the uncertainty of the interest rate for PWLB issued loans was analysed. Loans are available at different rates but all are linked to the interest rate of UK government bonds \cite{hmtreasuryConsultationIntroducingLocal2017}. The average interest rates for each month and for bonds with different maturity dates since 1998 are published by the DMO \cite{debtmanagementofficeHistoricalAverageDailyn.d.}. \newline For the calculation of the uncertainty range, the monthly interest rates of long-term UK government bonds published for 1998 up to 2020 were aggregated to annual averages. The actual and forecast annual interest rates were then compared across 10 year time horizon as summarised in Table \ref{tbl:InterestRate}. For a conservative estimate, the final uncertainty range of $R_{r}^{\%,\emph{min}}=-81\%$ and $R_{r}^{\%,\emph{max}}=0\%$ was determined based on the extreme value of $\Delta_{\%}$ in Table \ref{tbl:InterestRate}. 
\begin{table}[H]
\caption{Comparison of the forecast and actual interest rates of long-term UK government bonds across a 10 year time horizon. Based on annual averages, the actual interest rates between 2008 and 2020 are consistently overestimated if interest rates from 1998 until 2010 are used as forecasts.}
\label{tbl:InterestRate}
\centering
\begin{tabular}{c|c|c|c|c}
\toprule
\multicolumn{2}{c|}{Forecast} & \multicolumn{2}{c|}{Actual} & \multirow{2}{*}{$\Delta_{\%}$ [$\%$]} \\\cline{1-4}
Year & Interest rate [\%] & Year & Interest rate [\%] & \\\hline
1998 & 5.10 & 2008 & 4.43 & -13 \\
1999 & 4.58 & 2009 & 4.30 & -6 \\
2000 & 4.48 & 2010 & 4.30 & -4 \\
2001 & 4.60 & 2011 & 3.95 & -14 \\
2002 & 4.68 & 2012 & 3.12 & -33 \\
2003 & 4.60 & 2013 & 3.42 & -26 \\
2004 & 4.66 & 2014 & 3.25 & -30 \\
2005 & 4.34 & 2015 & 2.53 & -42 \\
2006 & 4.11 & 2016 & 1.99 & -52 \\
2007 & 4.49 & 2017 & 1.83 & -59 \\
2008 & 4.42 & 2018 & 1.84 & -58 \\
2009 & 4.30 & 2019 & 1.42 & -67 \\
2010 & 4.30 & 2020 & 0.81 & -81 \\
\bottomrule
\end{tabular}
\end{table}
Given the decline in the interest rates of long-term UK government bonds between 1998 and 2020 shown in Fig. \ref{fgr:InterestRate}, the actual interest rates are consistently overestimated when using historical interest rates as forecast. In the HIT model, a nominal discount rate of 7\% was assumed \cite{JALILVEGA20181051}. Hence, \emph{r} is varied between 1.33\% ($R_{r}^{\%,\emph{min}}=-81\%$) and 7\% ($R_{r}^{\%,\emph{max}}=0\%$) in the GSA which covers the range of interest rates observed between 1998 and 2019.
\subsection{Efficiency and capital cost of technologies}
The uncertainty ranges for the efficiencies and capital cost of the heating technologies and solar PV panels were determined based on the data catalogue published by \citet{andreas_v_olympios_2021_4692649}. In the catalogue, manufacturer datasheets and pricelists are compiled. \newline In \cite{andreas_v_olympios_2021_4692649}, specific capital costs (\emph{SC}) were tabulated and used to determine the uncertainty ranges of the capital cost. If the SC were observed to be a function of the name plate capacity (\emph{NPC}), a power law $\emph{SC}=a \cdot \emph{NPC}^b$ according to the economies of scale was fitted to the data points. The \emph{economies of scale} describe a concept in engineering according to which the cost of a piece of equipment and its capacity are not directly proportional because the cost correlates with the area but the capacity with the volume of the equipment \cite{phungTheoryEvidenceUsing1987}.\newline For air-sourced and ground-source heat pumps, a correlation between the coefficient of performance and the inlet temperature could be observed. Here, an exponential function $\emph{COP}(T_\emph{in}) = a \cdot \exp(b\cdot T_\emph{in})$ was fitted to the data points. For combined heat and power plants, the electrical efficiency was observed to be a function of \emph{NPC} and a power law $\eta_\emph{el}=a \cdot \emph{NPC}^b$ was fitted.\newline Whenever a fit function was applied, the uncertainty range was determined based on the deviation between the actual data points reported by the manufacturers and the fit function. However, if no fit function was applied, the uncertainty ranges were based on the deviation with respect to the average SC and performance over all data points. To differentiate between the use of a fit function or the mean value, a different representation was chosen for the Figures shown in Sections \ref{Sct:AppASHP} to \ref{Sct:AppBoiler}. If a fit function was applied, a dashed red line is used whereas a dashed orange line indicates the mean value across all data points. For all figures, the final bounds of the uncertainty range are represented by blue dashed lines and the data points considered in the analysis are shown as filled blue circles.\newline For the calculation of the uncertainty ranges, the data catalogue of \citet{andreas_v_olympios_2021_4692649} has two main limitations. While the specific capital costs include the value-added tax (VAT), the installation costs are excluded. As shown in the technical reports by BEIS, the installation cost however amount to 46\% for DH systems \cite{departmentofenergyclimatechangeAssessmentCostsPerformance2015} and 53\% for ASHP \cite{departmentforbusinessenergyindustrialstrategyCostInstallingHeating2020}. Therefore, the installation cost would be expected to increase the capital cost and its uncertainty range as they are strongly dependent on the specific project and the properties of the building \cite{departmentofenergyclimatechangeAssessmentCostsPerformance2015,departmentforbusinessenergyindustrialstrategyCostInstallingHeating2020}. The efficiency reported in the manufacturer data sheets refers to the technical unit itself under standardized, laboratory conditions. Any variation in performance of the technologies that might be observed in the real energy system because of incorrect operation, differences in the buildings' heating system design or thermal efficiency is not taken into account. Extensive data collected in field trials for all types of technologies were not available. Therefore, the uncertainty ranges derived based on \cite{andreas_v_olympios_2021_4692649} should be seen as lower bounds on the uncertainty of the real performance.

\subsubsection{Air-sourced heat pumps}\label{Sct:AppASHP}
\begin{figure}[H]
\centering
\includegraphics[width=0.9\textwidth]{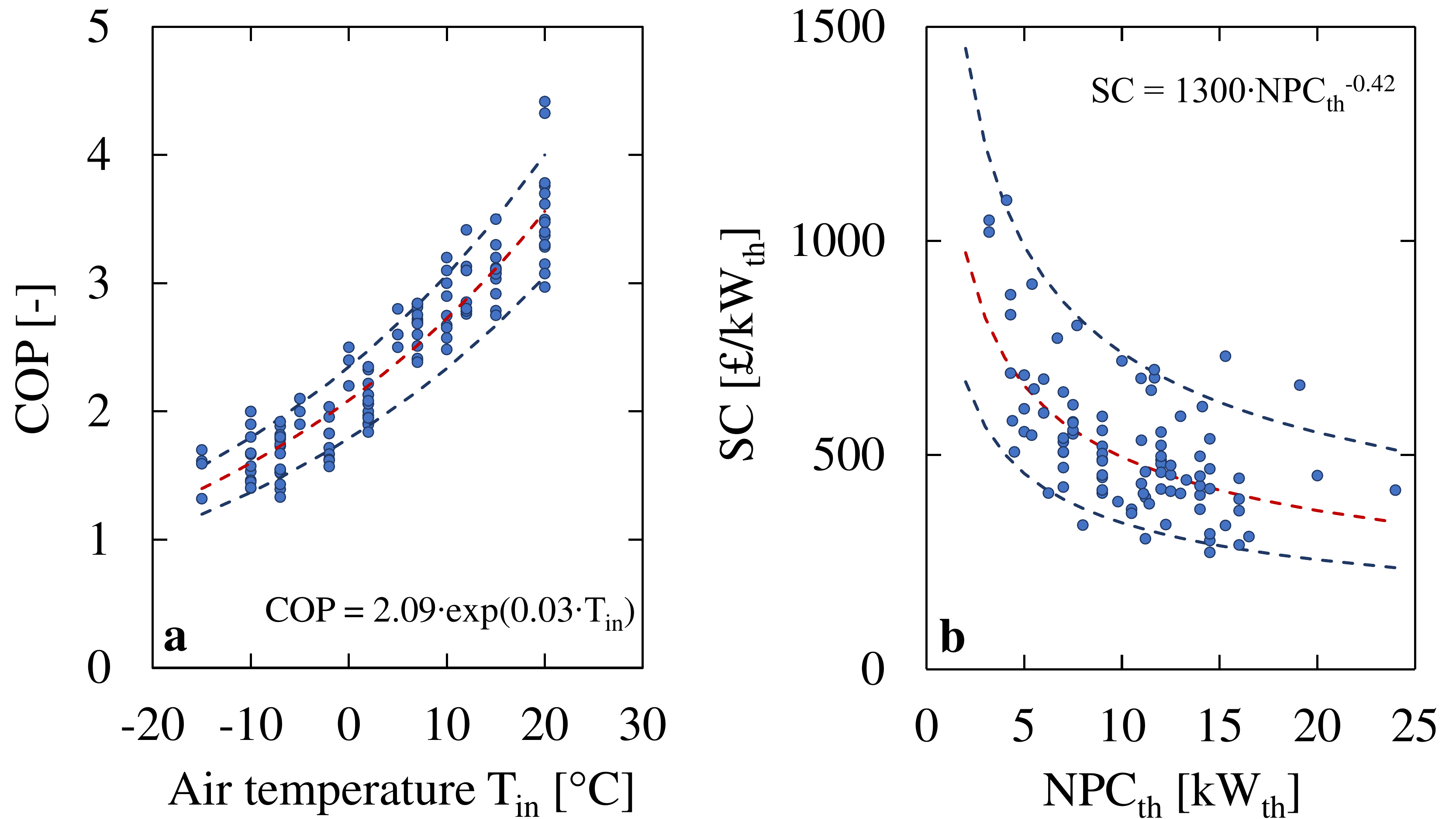}
\caption{In (a), the coefficient of performance (\emph{COP}) is shown to increase with the outside air temperature ($T_\emph{in}$) in air-water heat pumps for a fixed outlet temperature of $55~^\circ$C. Lower specific costs (\emph{SC}) for higher heat output capacities ($\emph{NPC}_\emph{th}$) are observed in (b). For each data set (\emph{filled blue circles}), the uncertainty ranges (\emph{dashed blue lines}) were determined with respect to the fit function (\emph{dashed red line}).}
\label{fgr:ASHP}
\end{figure}

In \cite{andreas_v_olympios_2021_4692649}, the \emph{COP} and \emph{NPC}$_\emph{th}$ of air-water ASHP for outlet temperatures $T_\emph{out}$ of 35~$^\circ$C, 45~$^\circ$C and 55~$^\circ$C are tabulated. For this analysis, the data set for $T_\emph{out}=55~^\circ$C was used as an outlet temperature of about 50 $^\circ$C is required to use air-source heat pumps for domestic hot water (DHW) and space heating in the current British housing stock \cite{departmentforbusinessenergyindustrialstrategyEvidenceGatheringLow2016,deltaenergyenvironmentIEAHPPAnnex2014}. The analysis was further limited to the ASHP units using R410a as a refrigerant\footnote{The R410a refrigerant is based on a mixture of difluoromethane and pentafluoroethane. It will be phased out due to its high Greenhouse Warming Potential \cite{XU2013892}.} because the data set was more comprehensive. \newline For 25 ASHP units, the \emph{COP} was reported for different outdoor air temperature $T_\emph{in}$. A correlation of \emph{COP} and the outdoor air temperature $T_\emph{in}$ was identified and is shown in Fig. \ref{fgr:ASHP}a. Therefore, an exponential function $\emph{COP}(T_\emph{in}) = a \cdot \exp(b\cdot T_\emph{in})$ was fitted to the data first. Afterwards, the largest deviation of the actual $\emph{COP}(T_\emph{in})$ published by the manufacturer from the $\emph{COP}(T_\emph{in})$ predicted by the fit function was determined for each $T_\emph{in}$. As a result, a conservative estimate for the uncertainty range at each $T_\emph{in}$ was obtained. For the final uncertainty range of $R^{\%,\emph{min}}_{COP_{ASHP}}=-14\%$ and $R^{\%,\emph{max}}_{COP_{ASHP}}=12\%$, the upper and lower bounds of the ranges for each $T_\emph{in}$ were averaged. \emph{COP}(-15~$^\circ$C) and \emph{COP}(20~$^\circ$C) were excluded because the annual outdoor temperature in Great Britain rarely drops below -10~$^{\circ}$C \cite{PFENNINGER20161251} and the heating degree days are defined for outdoor temperatures below 15.5~$^{\circ}$C for the UK \cite{macleayLongTermMean2013}.
For 101 ASHP units, the specific cost could be calculated based on the capital cost and the nameplate capacity $\emph{NPC}_\emph{th}$ for the heat output listed in \cite{andreas_v_olympios_2021_4692649}. As \emph{SC} was observed to be dependent on $\emph{NPC}_\emph{th}$ (see Fig. \ref{fgr:ASHP}b), the power law 
\begin{equation}\label{eq:SCASHP}
    SC = 1300\cdot NPC_{th}^{-0.42}
\end{equation}
was fitted to the data points. To characterise the uncertainty around the fit function, the capacity range 3.2~kW$_\text{th}$~$\le$~$\emph{NPC}_\emph{th}$~$\le$~24 kW$_\text{th}$ was first divided into 2~kW$_\text{th}$ intervals. Within each interval, the SC predicted by Eq. \ref{eq:SCASHP} was compared to the actual manufacturer data. Based on the largest deviations from the fit function, a conservative uncertainty range for each capacity interval was determined. The final relative uncertainty range of $R^{\%,\emph{min}}_{Cost^C_{ASHP}}=-31\%$ and $R^{\%,\emph{max}}_{Cost^C_{ASHP}}=49\%$ was calculated by averaging the upper and lower bounds across all intervals, respectively.

\subsubsection{Ground-source heat pumps}\label{Sct:AppGSHP}
\begin{figure}[H]
\centering
  \includegraphics[width=0.9\textwidth]{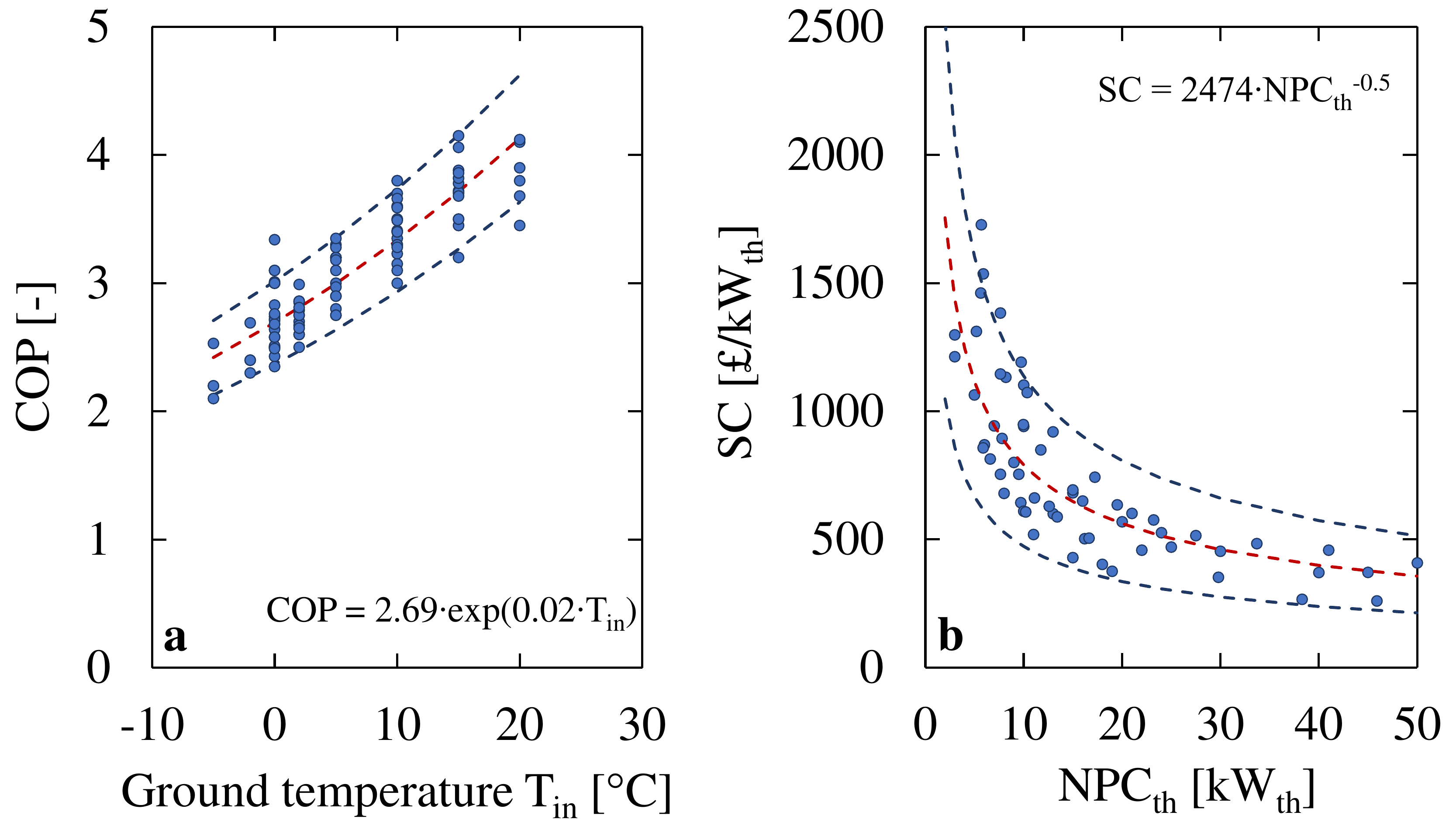}
  \caption{For ground-source heat pumps, (a) the coefficient of performance (\emph{COP}) increases with the ground temperature ($T_\emph{in}$) and (b) the specific costs (\emph{SC}) decrease with increasing heat output capacities ($\emph{NPC}_\emph{th}$). In (a) and (b), the uncertainty ranges (\emph{dashed blue lines}) are based on the deviation of the manufacturer data (\emph{filled blue circles}) from the fit function (\emph{dashed red lines}).}
  \label{fgr:GSHP}
\end{figure}

Similar to ASHP, the COP of  brine-water GSHP is dependent on the ground temperature $T_\emph{in}$ (see Fig. \ref{fgr:GSHP}a). For 31 different brine-water GSHP units, the COP between 0~$^{\circ}$C~$\le$~$T_\emph{in}$~$\le$~15~$^{\circ}$C for an outlet temperature $T_\emph{out}=55~^\circ$C are reported. Consistent with the methodology applied to ASHP, an exponential function was fitted to all data points first. Horizontal ground loops are installed at a minimum depth of 1~m below the ground \cite{doi:https://doi.org/10.1002/9781118447512.ch11} where the soil temperature varies between 2-13 $^{\circ}$C \cite{metofficeMIDASUKSoil2006} during the heating months (January-April, October-December) \cite{buildingresearchestablishmentEnergyFollowupSurvey2013}. Therefore, $T_\emph{in}$=15~$^\circ$C and $T_\emph{in}$=0~$^\circ$C were excluded from the further analysis. A conservative uncertainty range was calculated for each $T_\emph{in}$ by comparing the predicted $\emph{COP}(T_\emph{in})$ to the maximum and minimum $\emph{COP}(T_\emph{in})$ value reported by the manufacturers, respectively. Finally, the individual uncertainty ranges were averaged to give the uncertainty range of $R^{\%,\emph{min}}_{COP_\emph{GSHP}}=-12\%$ and $R^{\%,\emph{max}}_{COP_\emph{GSHP}}=12\%$.   
For the calculation of the specific costs, the tabulated heat output capacities were divided by the capital costs of the units. In \cite{andreas_v_olympios_2021_4692649}, only the heat output capacity $\emph{NPC}_\emph{th}$ at an outlet temperature of $T_\emph{out}=35~^\circ$C is reported. Smaller heat output capacities are expected for outlet temperatures $T_\emph{out}\ge$50~$^\circ$C that are required for the British housing stock. Given the lack of data, it was assumed, that the uncertainty ranges determined herein would also be applicable to the specific costs determined based on the $\emph{NPC}_\emph{th}$ at $T_\emph{out}\ge$50~$^\circ$C.\newline For the 66 GSHP units shown in Fig. \ref{fgr:GSHP}b, the SC decrease with increasing $\emph{NPC}_\emph{th}$ and similary to ASHP the dependence is described by 
\begin{equation}\label{eq:SCGSHP}
    SC = 2474\cdot NPC_{th}^{-0.5}.
\end{equation}
For the uncertainty characterisation, the full capacity range 3~kW$_\text{th}$~$\le$~$\emph{NPC}_\emph{th}$~$\le$~30 kW$_\text{th}$ was first divided into 10~kW$_\text{th}$ intervals. Within each interval, the $SC$ function in Eq. \ref{eq:SCASHP} was linearly approximated using its average value that was subsequently compared to the actual manufacturer data. Based on the largest deviations, a conservative uncertainty range was determined for each capacity interval. The final relative uncertainty range of $R^{\%,\emph{min}}_{Cost^C_{GSHP}}=-40\%$ and $R^{\%,\emph{max}}_{Cost^C_{GSHP}}=44\%$ was calculated by averaging the upper and lower bounds across all intervals, respectively.\newline
As Eq. \ref{eq:SCASHP} and \ref{eq:SCGSHP} highlight, brine-water GSHPs generally tend to be more expensive than air-water HP. The performances of air-water HPs and brine-water GSHPs both increase with increasing inlet temperatures as less work by the compressor is required to reach the same outlet temperature. However, as the comparison of Fig. \ref{fgr:ASHP}a and Fig. \ref{fgr:GSHP}a shows, the COP of brine-water GSHPs increases at a slower rate. While the heat capacity of air increases slightly between -15 $^{\circ}$C and 20 $^{\circ}$C, the heat capacity of water decreases. Therefore, less heat can be transferred from the water to the refrigerant, which reduces the rate by which the performance increases. 

\subsubsection{Combined heat and power plants}\label{Sct:AppCHP}
\begin{figure}[H]
\centering
  \includegraphics[width=0.9\textwidth]{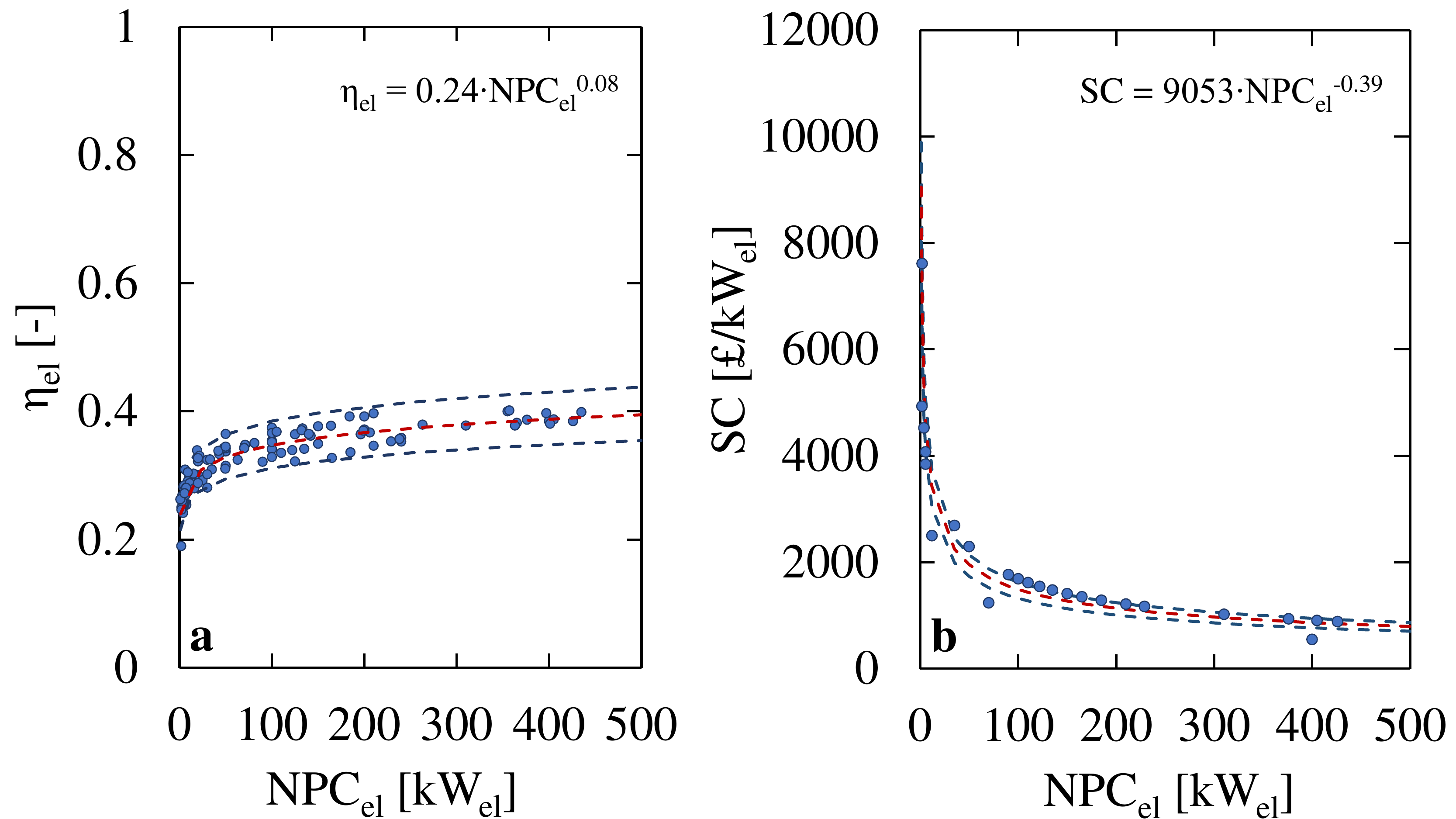}
  \caption{For the 123 combined heat and power (CHP) plants shown in (a), the electrical efficiency ($\eta_\emph{el}$) and the respective electrical output capacity ($\emph{NPC}_{\emph{el}}$) was provided in \cite{andreas_v_olympios_2021_4692649}. However, for only 41 units, also the specific costs (\emph{SC}) were available. For illustration purposes only 24 are shown in (b). A power law (\emph{dashed red line}) was fitted to both data sets (\emph{filled blue circles}), respectively. The uncertainty ranges (\emph{dashed blue line}) are significantly smaller for SC than $\eta_\emph{el}$ as the few data points are spread over a large capacity interval.}
  \label{fgr:CHP}
\end{figure}

\begin{figure}[H]
\centering
  \includegraphics[width=0.9\textwidth]{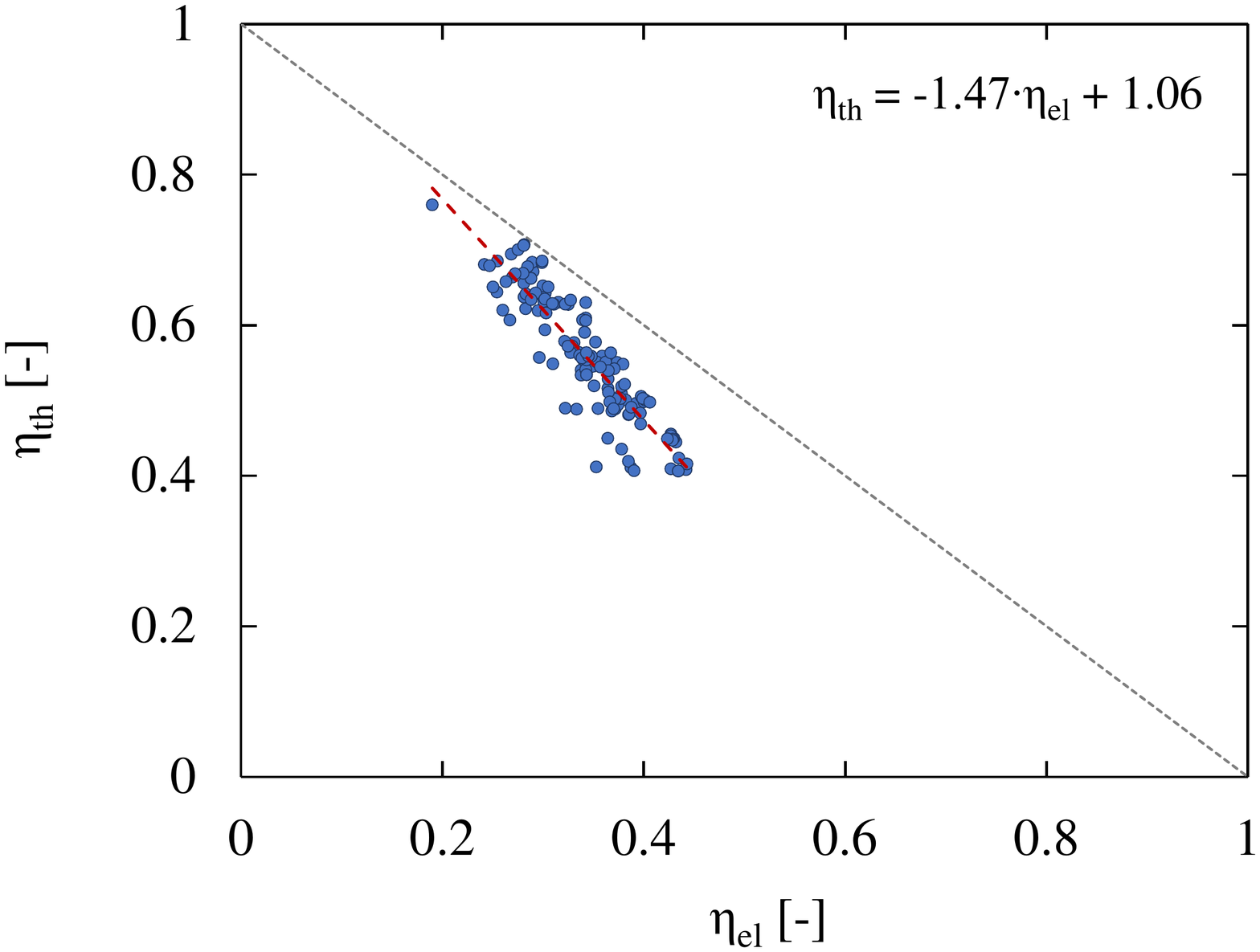}
  \caption{According to thermodynamics, the sum of the electrical ($\eta_\emph{el}$) and thermal ($\eta_\emph{th}$) efficiencies of combined heat and power (CHP) plants must be smaller than 1 (\emph{dashed grey line}). The dependency of $\eta_\emph{th}$ on $\eta_\emph{el}$ can be expressed by a linear relationship (\emph{dashed red line}) that holds for the range of $0.19 \le \eta_\emph{el} \le 0.44$.}
  \label{fgr:CHPEfficiencies}
\end{figure}

Overall, 140 internal combustion engine combined heat and power plants (ICE-CHP) fuelled by natural gas with electrical output capacities between 1 kW$_\text{el}$ $\le\emph{NPC}_\emph{el}\le$ 9020~kW$_\text{el}$ are provided in \cite{andreas_v_olympios_2021_4692649}. The 123 ICE-CHP units within 1 kW$_\text{el}$ $\le \emph{NPC}_\emph{el}\le$ 450~kW$_\text{el}$ were considered in the uncertainty characterisation for the electrical efficiency $\eta_\emph{el}$. 17 ICE-CHP units with 450 kW$_\text{el}<$ $\emph{NPC}_\emph{el}$ were excluded as they were sparsely distributed across the large capacity interval. As shown in Fig. \ref{fgr:CHP}a, $\eta_\emph{el}$ is a function of $\emph{NPC}_\emph{el}$ that can be described by 
\begin{equation}
    \eta_\emph{el} = 0.24 \cdot \emph{NPC}_\emph{el}^{0.08}.
\end{equation}
The increase in $\eta_\emph{el}$ with increasing capacity is explained by the reduction in the surface-to-volume ratio in the combustion engine. For small scale ICE-CHP units, the surface-to-volume ratio is higher and hence a larger share of the gas is in contact with the cylinder walls \cite{brownInvestigationScalingLaws2016} \cite{MARUTA2011125}. As the surface-to-volume ratio decreases with larger capacities, larger volumes can be combusted at similar thermal losses. However, $\eta_\emph{el}$ reaches a plateau at around $\eta_\emph{el}=$37\% for units larger than 100 kW$_\text{el}$. 
For the calculation of the uncertainty ranges, the full $\emph{NPC}_\emph{el}$ range was divided into intervals in which the fit function was approximated by its average. The maximum and minimum $\eta_\emph{el}$ reported by the manufacturers were used to calculate the relative deviations from the fitted average $\eta_\emph{el}$ for each interval. The final range $R^{\%,\emph{min}}_{\eta_\emph{el}}=-10\%$ and $R^{\%,\emph{max}}_{\eta_\emph{el}}=11\%$ was determined by averaging the upper and lower bounds across all intervals, respectively.\newline Dictated by the first and second law of thermodynamics, the overall efficiency of a ICE-CHP unit is smaller than 1  \cite{schmidtTechnicalThermodynamicsEngineers2019}. Hence, the electrical and thermal efficiency are interdependent. The more thermal energy of the combustion is converted to electrical energy, the smaller the thermal efficiency. Figure \ref{fgr:CHPEfficiencies} shows that all data points are positioned in the region of thermodynamic feasibility and that the relationship can be approximated by a line described as 
\begin{equation}
    \eta_\emph{th} = -1.47 \cdot \eta_\emph{el}+1.06.
\end{equation}
While the electricity is generally generated from the piston driving the crankshaft mechanism of a grid-synchronised AC generator \cite{doi:10.1680/ener.2010.163.4.143}, the heat sources from a gas engine CHP are more varied including the jacket water cooler, the exhaust gas cooler, the intercooler and the lubrication oil cooler \cite{MARTENS1998935,departmentforbusinessenergyindustrialstrategyCombinedHeatPower2021}. Hence, more design options allow for more variance in the thermal efficiencies and therefore larger absolute uncertainty ranges. However, in relative terms they are comparable to the uncertainty ranges of $\eta_\emph{el}$ as shown in the Supplementary Material.\newline Note that the efficiencies discussed here refer to the operation of the ICE-CHP units at full load. Based on additional data provided in \citet{andreas_v_olympios_2021_4692649}, the electrical efficiency decreases on average by 5\% (at 75\% load) and 13\% (at 50\% load) in part-load operation.\newline For only 41 out of 140 tabulated ICE-CHP units, the capital cost are available in \cite{andreas_v_olympios_2021_4692649} and thereby allowed for the calculation of the specific costs. The few data points are spread across the full capacity interval, 1 kW$_\text{el}$ $\le\emph{NPC}_\emph{el}\le$ 9020~kW$_\text{el}$. Because of the limited data availability, no cut-off at 450~kW$_\text{el}$ was applied but the full capacity range was considered. As shown in Fig. \ref{fgr:CHP}b, also here the economies of scale could be applied:
\begin{equation}\label{eq:CHPcost}
    SC = 9053\cdot NPC_{el}^{-0.4}
\end{equation}
For each data point, the actual \emph{SC} published by the manufacturer was compared to the \emph{SC} predicted by \ref{eq:CHPcost}. The averaging of the upper and lower bounds of each relative uncertainty range resulted in the final uncertainty range of $R^{\%,\emph{min}}_{\emph{Cost}^C_\emph{CHP}}=-11\%$ and $R^{\%,\emph{max}}_{\emph{Cost}^C_\emph{CHP}}=9\%$.
\subsubsection{Solar photovoltaic}\label{Sct:AppPV} 
\begin{figure}[H]
\centering
  \includegraphics[width=0.9\textwidth]{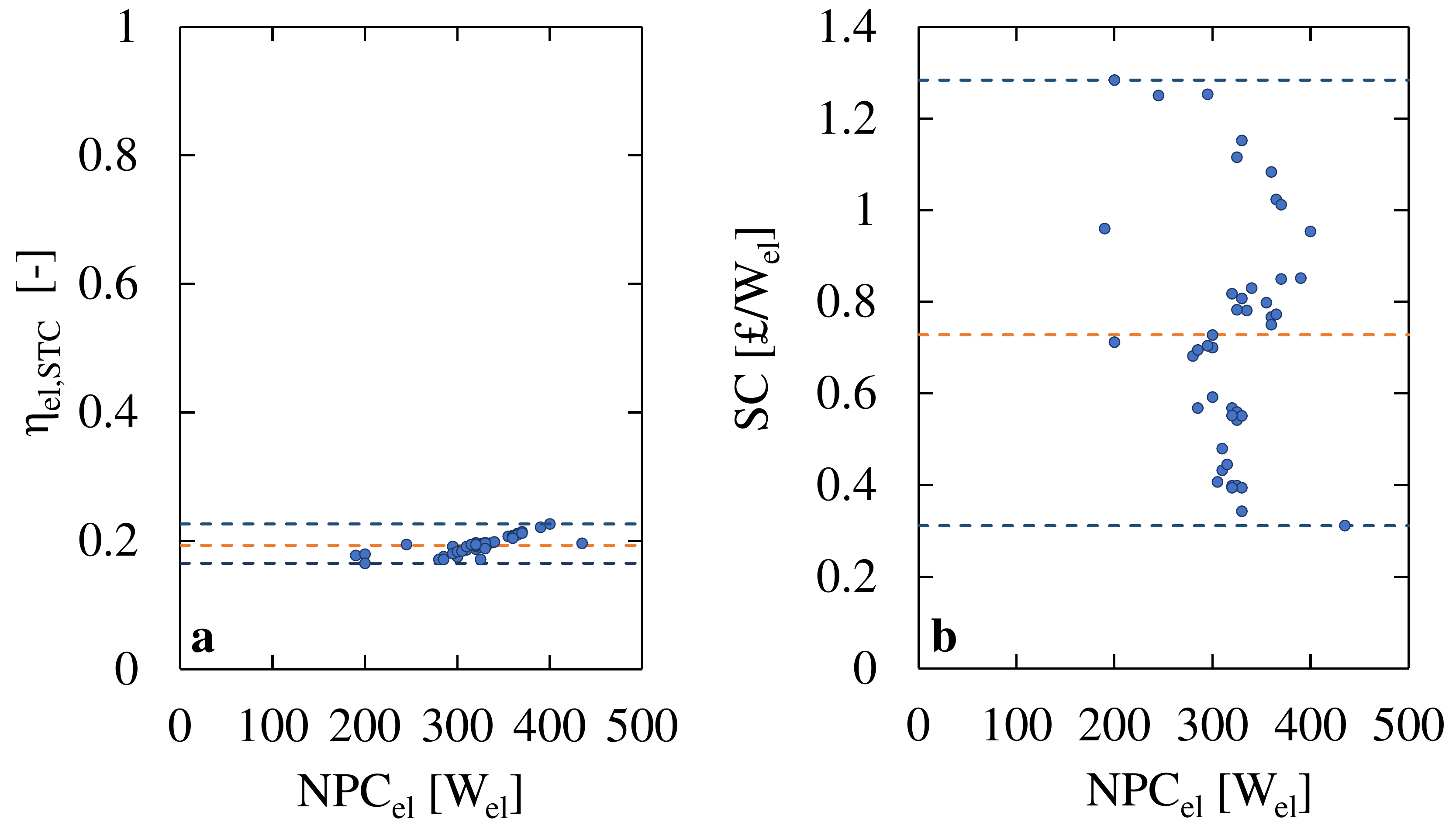}
  \caption{For 44 monocrystalline silicone photovoltaic panels (\emph{filled blue circles}), (a) the efficiencies under standard test conditions ($\eta_\emph{el,STC}$) and (b) the specific costs (\emph{SC}) with respect to the nominal power ($\emph{NPC}_{\emph{el}}$) are shown. The average $\eta_\emph{el,STC}$ and \emph{SC}, respectively, are used as a reference value (\emph{dashed orange lines}) to determine the uncertainty ranges (\emph{dashed blue lines}).}
  \label{fgr:PV}
\end{figure}

The data provided for monocrystalline silicone PV panels in \cite{andreas_v_olympios_2021_4692649} are based on the German market. The prices are converted using an exchange rate of 0.89 GBP/EUR, the annual average exchange rate in 2020 \cite{officefornationalstatisticsAverageSterlingExchange2022}.\newline For 44 solar PV panels, the efficiency under standard test conditions (STC)\footnote{Standard test conditions (STC) are defined as a cell temperature of 25~$^{\circ}$C, solar radiation of 1 kW/m$^2$, and no wind.}, $\eta_\emph{el,STC}$, is tabulated in \cite{andreas_v_olympios_2021_4692649} and shown in Fig. \ref{fgr:PV}a. A linear relationship for nominal powers 300~W$_\text{el}\le\emph{NPC}_\emph{el}\le$~400 W$_\text{el}$ is observed. However, as this correlation could not confirmed in the literature, it was disregarded. Instead, the average $\eta_\emph{el,STC}$ over all data points was chosen as a reference value. As a conservative estimate, the final uncertainty range of $R^{\%,\emph{min}}_{\eta_\emph{PV}}=-14\%$ and $R^{\%,\emph{max}}_{\eta_{\emph{PV}}}=17\%$ was determined based on the relative deviation of the maximum and minimum $\eta_\emph{el,STC}$ reported by the manufacturers from the reference value. \newline The real efficiencies of solar PV panel installations are generally expected to be smaller than $\eta_\emph{el,STC}$, as $\eta_\emph{el,STC}$ only refers to the efficiency of the solar PV panel itself, neglecting any efficiency losses introduced by the inverter (2-6\%) and output power line (1\%) \cite{doi:https://doi.org/10.1002/9781118927496.ch47}. Additionally, higher uncertainties for the installations' performances are expected because the efficiency also strongly depends on the angle-of-incidence, irradiance intensity and module temperature \cite{doi:10.1063/1.5128171}.\newline As Figure \ref{fgr:PV}b shows, the specific cost are observed to be independent of the nominal power of the PV module. Similar to the procedure applied for $\eta_\emph{el,STC}$, the specific costs \emph{SC} were first averaged across all capacities. Then, the conservative uncertainty range of $R^{\%,\emph{min}}_{\emph{Cost}^C_\emph{PV}}=-57\%$ and $R^{\%,\emph{max}}_{\emph{Cost}^C_\emph{PV}}=76\%$ was determined as the relative deviation of the maximum and minimum \emph{SC} provided by the manufacturers from the  reference value, respectively.
\subsubsection{Natural gas boiler}\label{Sct:AppBoiler}
\begin{figure}[H]
\centering
  \includegraphics[width=0.9\textwidth]{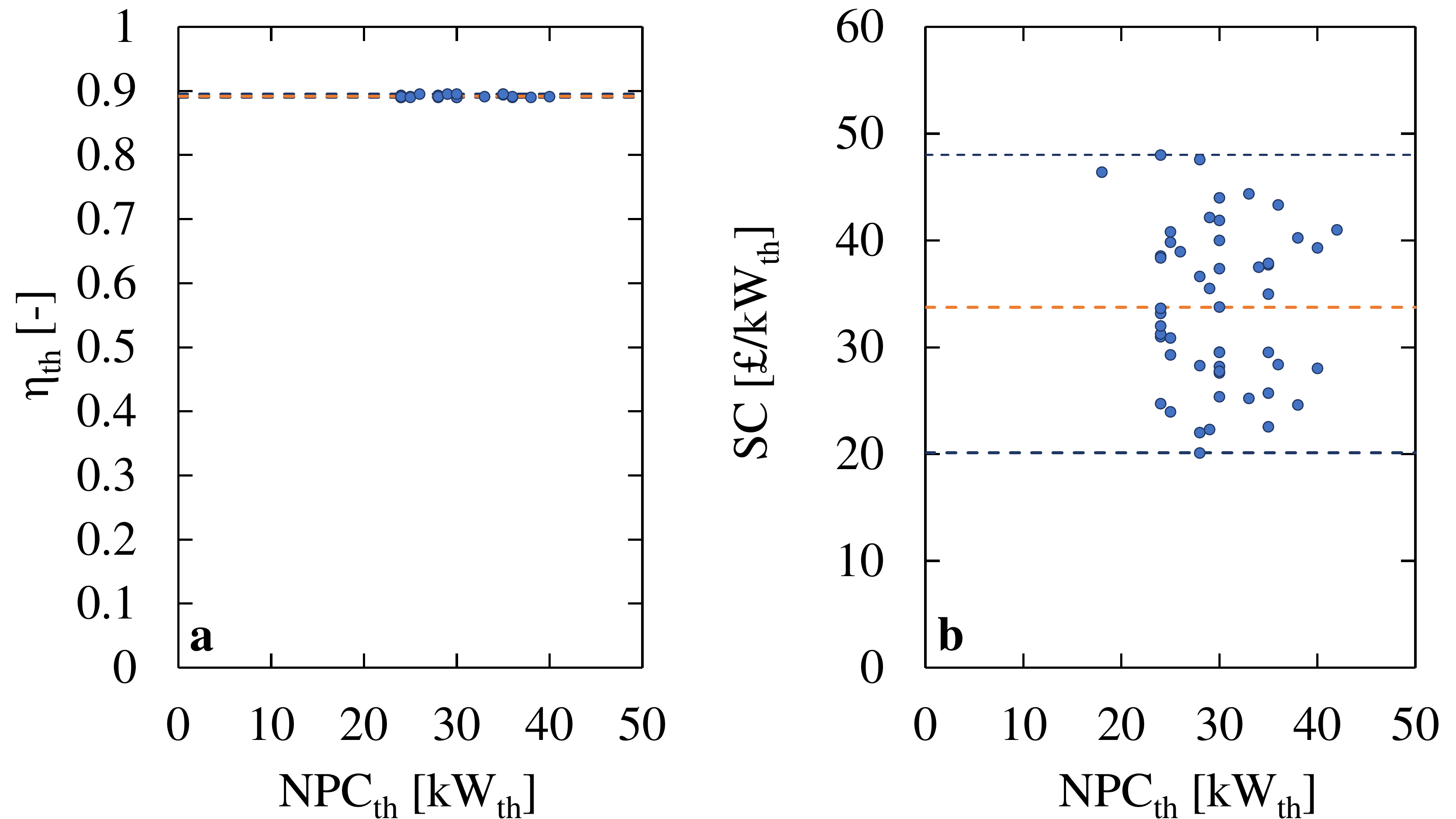}
  \caption{The efficiency ($\eta_\emph{th}$) and specific costs (\emph{SC}) appear to be independent of the nameplate capacity (\emph{NPC}$_\emph{th}$) of natural gas boilers. Not all 23 units (\emph{filled blue circles}) in (a) are visible as many overlap given the small range of $89\% \le \eta_\emph{th} \le 89.5\%$. For a given \emph{NPC}$_\emph{th}$, \emph{SC} however varies significantly. The uncertainty ranges (\emph{dashed blue lines}) are calculated with respect to the average SC across 41 units (\emph{dashed orange lines}). }
  \label{fgr:NGBoiler}
\end{figure}

The HIT model described in Section \ref{Sct:HITmodel} includes natural gas boilers as heating technologies for individual households as well as for the supply of district heating. While \cite{andreas_v_olympios_2021_4692649} contains data on both domestic and large-scale natural gas boilers, the efficiencies for the 41 large-scale natural gas boilers with thermal output capacities ($\emph{NPC}_\emph{th}$) of 58~kW$_\text{th}\le\emph{NPC}_\emph{th}\le$ 2900~kW$_\text{th}$ are missing. Therefore, the data for domestic natural gas boilers with 18~kW$_\text{th}\le\emph{NPC}_\emph{th}\le$ 42~kW$_\text{th}$ were also used to characterise the uncertainties of DH boilers.\newline 
For 27 domestic natural gas boilers, the average efficiency for the provision of domestic hot water and space heating was tabulated in \cite{andreas_v_olympios_2021_4692649} and is shown in Fig. \ref{fgr:NGBoiler}a. As the efficiencies for all units lie within the narrow interval of $89\%\le\eta_\emph{th}\le89.5\%$, not all data points are visible due to their overlap. To determine the conservative uncertainty range, the efficiencies of the 41 data points were averaged first and subsequently used as a reference value. The relative upper and lower bound was calculated based on the deviation of the maximum and minimum efficiency published by the manufacturers from this reference value, respectively. As a result, the final uncertainty range of $R^{\%,\emph{min}}_{\eta_B}= -0.2\%$ and $R_{\eta_B}^{\%,\emph{max}}=0.4\%$ was identified.\newline The units tabulated in \cite{andreas_v_olympios_2021_4692649} can be identified as condensing boilers given their space heating efficiencies of 93-94\%. However, older, less efficient natural gas boiler are still in use as 26\% of English houses had no condensing boiler installed by 2019 \cite{ministryofhousingcommunitiesandlocalgovernmentEnglishHousingSurvey2021}. Therefore, the average efficiency for DHW and space heating shown in Fig. \ref{fgr:NGBoiler}a should be seen as an upper bound on the real efficiencies across the British housing stock. Additionally, the true uncertainty range for the performance of domestic natural gas boiler in the real energy system is likely to be larger than determined based on the data in \cite{andreas_v_olympios_2021_4692649}.\newline 
The thermal output capacities ($\emph{NPC}_\emph{th}$) and the capital costs for all 41 units are included in \cite{andreas_v_olympios_2021_4692649} and can therefore be used to determine the specific costs (\emph{SC}). All data points are shown in Fig. \ref{fgr:NGBoiler}b. The \emph{SC} of all units were averaged first, before comparing the mean value to the maximum and minimum SC tabulated in \cite{andreas_v_olympios_2021_4692649}. The relative deviation from the mean was used as a conservative estimate for the uncertainty range of $R^{\%,\emph{min}}_{\emph{Cost}^C_B}=-40\%$ and $R^{\%,\emph{max}}_{\emph{Cost}^C_B}=42\%$.

\pagebreak
\newpage

\bibliography{PublicationV1.bib}
\bibliographystyle{model3-num-names}

\end{document}

%% file: variables.tex
\nomenclature[A]{MILP}{Mixed integer linear programming}
\nomenclature[A]{UK}{United Kingdom}
\nomenclature[A]{US}{United States}
\nomenclature[A]{ESM}{Energy system model}
\nomenclature[A]{EU}{European Union}
\nomenclature[A]{GSA}{Global sensitivity analysis}
\nomenclature[A]{GHG}{Greenhouse gas}
\nomenclature[A]{GSHP}{Ground-source heat pump}
\nomenclature[A]{ASHP}{Air-source heat pump}
\nomenclature[A]{LSA}{Local sensitivity analysis}
\nomenclature[A]{B}{Natural gas boiler}
\nomenclature[A]{LA}{Local authority}
\nomenclature[A]{LSOA}{Lower level super output area}
\nomenclature[A]{MSOA}{Middle level super output area}
\nomenclature[A]{CHP}{Combined heat and power plant}
\nomenclature[A]{COP}{Coefficient of performance}
\nomenclature[A]{HEX}{Heat exchanger}
\nomenclature[A]{DH}{District heating}
\nomenclature[A]{HIT}{Heat Infrastructure and Technology}
\nomenclature[A]{EE}{Elementary Effect}
\nomenclature[A]{PV}{Photovoltaic}
\nomenclature[A]{Erad}{Electric resistance heater}
\nomenclature[A]{BEIS}{Department for Business, Energy \& Industrial Strategy}
\nomenclature[A]{UEP}{Updated energy and emission projections}
\nomenclature[A]{NPC}{Nameplate capacity}
\nomenclature[A]{NPV}{Net present value}
\nomenclature[A]{ICE}{Ignition combustion engine}
\nomenclature[A]{SC}{Specific cost}

\nomenclature[G]{$\boldsymbol{\mathrm{\theta}}$}{vector of uncertain input parameters}
\nomenclature[G]{$\varphi$}{modelling choice}
\nomenclature[G]{$\mu_{ij}$}{mean of the EE distribution for input factor $x_i$ on output variable $y_j$}
\nomenclature[G]{$\mu_{ij}^{*}$}{mean of the absolute EE distribution for input factor $x_i$ on output variable $y_j$}
\nomenclature[G]{$\mu_{ij}^{*,sc}$}{scaled mean of the absolute EE distribution for input factor $x_i$ on output variable $y_j$}
\nomenclature[G]{$\sigma_{ij}$}{standard deviation of the EE distributions for input factor $x_i$ on output variable $y_j$}
\nomenclature[G]{$\Delta_\phi$}{change to the modelling choice $\phi$ for each trajectory on the unit hypercube}
\nomenclature[G]{$\Delta_\theta$}{change to each input parameter $\theta_k$ along each trajectory on the unit hypercube}
\nomenclature[G]{$\eta$}{efficiency}
\nomenclature[G]{$\boldsymbol{\mathrm{\gamma}}$}{vector of the sample points within the input factor space}
\nomenclature[G]{$\delta y_{ji}$}{change in the output variable $y_j$ upon a change in the input factor $x_i$}
\nomenclature[G]{$\Tilde{\sigma}_{j}$}{standard deviation of the $y_j$ distribution for all sample points $G$}

\nomenclature[V]{$f$}{uncertainty distribution}
\nomenclature[V]{$K$}{total number of uncertain input parameters}
\nomenclature[V]{$r$}{discount rate}
\nomenclature[V]{$I$}{total number of input factors}
\nomenclature[V]{$T$}{total number of trajectories}
\nomenclature[V]{$t$}{trajectory}
\nomenclature[V]{$G$}{total number of sample points on the unit hypercube}
\nomenclature[V]{$\mathbf{x}$}{vector of input factors}
\nomenclature[V]{$\mathbf{y}$}{vector of output variables}
\nomenclature[V]{$R^{min}$}{Lower bound of the absolute uncertainty range}
\nomenclature[V]{$R^{\%,min}$}{Lower bound of the relative uncertainty range}
\nomenclature[V]{$R^{max}$}{Upper bound of the absolute uncertainty range}
\nomenclature[V]{$R^{\%,max}$}{Upper bound of the relative uncertainty range}
\nomenclature[V]{$p_\theta$}{number of levels for input parameters $\theta$ on unit hypercube}
\nomenclature[V]{$p_\varphi$}{number of levels for the modelling choice $\phi$ on a unit hypercube}

\nomenclature[I]{$i$}{type of input factor}
\nomenclature[I]{$j$}{type of output variable}
\nomenclature[I]{$k$}{type of input parameter}
\nomenclature[I]{$g$}{type of sample point}

\nomenclature[B]{$E$}{electricity}
\nomenclature[B]{$H$}{heat}
\nomenclature[B]{$G$}{natural gas}
\nomenclature[B]{$\emph{spres}$}{spatial resolution}
\nomenclature[B]{$\emph{tech}$}{all technologies (ASHP, GSHP, PV, B, Erad, HEX)}

\nomenclature[P]{$\emph{CO2}$}{carbon dioxide}
\nomenclature[P]{$\emph{Th}$}{thermal}
\nomenclature[P]{$\emph{E}$}{electricity}
\nomenclature[P]{$C$}{capital}
\nomenclature[P]{$M$}{operating \& maintenance}
\nomenclature[P]{$\emph{ND}$}{intranodal network}
\nomenclature[P]{$\emph{NT}$}{internodal network}

\nomenclature[H]{$\emph{TOTALCOSTS}$}{net present value of the total system cost}
\nomenclature[H]{$\emph{CPT}$}{net present value of the total system's capital cost}
\nomenclature[H]{$\emph{FE}$}{net present value of the total system's gas and electricity cost}
\nomenclature[H]{$\emph{CO}_2$}{total carbon dioxide emissions}
\nomenclature[H]{$\emph{NCH}_{\emph{tech}}$}{newly installed capacity of technology}
\nomenclature[H]{$\emph{NLN}_{E/G/H}$}{new network length of electricity, gas or heat networks installed within a node}
\nomenclature[H]{$\emph{ICN}_{E/G/H}$}{new capacity of electricity, gas or heat networks installed between two nodes}
\nomenclature[H]{$\emph{ELEC}_{\emph{TOT}}$}{total electricity consumed from the grid}
\nomenclature[H]{$\emph{FUEL}_{\emph{TOT}}$}{total natural gas consumed}